\documentclass[%
 a4paper,
 reprint,
superscriptaddress,
 amsmath,amssymb, amsfonts,
 aps,longbibliography, prl%prx
]{revtex4-2}

\usepackage{graphicx}% Include figure files
\usepackage{dcolumn}% Align table columns on decimal point
\usepackage{color}
\usepackage{bm}% bold math
\usepackage{hyperref}% add hypertext capabilities
\usepackage{lipsum}
\usepackage[caption = false]{subfig}

\newcommand{\bra}[1]{\langle #1|}
\newcommand{\ket}[1]{| #1 \rangle}
\newcommand{\op}[1]{\hat{#1}}
\newcommand{\mc}[1]{\mathcal{#1}}
\newcommand{\cre}[1]{\hat{#1}^\dagger}
\newcommand{\des}[1]{\hat{#1}}
\newcommand{\Id}{\mathrm{Id}}
\newcommand{\Tr}{\mathrm{Tr}}
\newcommand{\proj}[1]{\ket{#1}\bra{#1}}
\newcommand{\erw}[1]{\langle #1 \rangle}
\newcommand{\abs}[1]{\left| #1 \right|}
\newcommand{\bkt}[2]{\bra{#1} #2 \rangle}

\begin{document}

\preprint{}

\title{Quantum Fluctuation Theorem for Arbitrary Measurement and Feedback Schemes}

\author{Kacper Prech}
\email{kacper.prech@unibas.ch}
\author{Patrick P. Potts}
\affiliation{Department of Physics, University of Basel, Klingelbergstrasse 82, 4056 Basel, Switzerland}


\begin{abstract}

Fluctuation theorems and the second law of thermodynamics are powerful relations constraining the behavior of out-of-equilibrium systems. While there exist generalizations of these relations to feedback controlled quantum systems, their applicability is limited, in particular when considering strong and continuous measurements. In this Letter, we overcome this shortcoming by deriving a novel fluctuation theorem, and the associated second law of information thermodynamics, which remain applicable in arbitrary feedback control scenarios. In our second law, the entropy production is bounded by the coarse-grained entropy production that is inferrable from the measurement outcomes, an experimentally accessible quantity that does not diverge even under strong continuous measurements. We illustrate our results by a qubit undergoing discrete and continuous measurement, where our approach provides a useful bound on the entropy production for all measurement strengths.

\end{abstract}

\maketitle

\emph{Introduction--} 
Most systems of relevance, ranging from technological devices to living organisms, are out of equilibrium.
Stochastic thermodynamics~\cite{Seifert_2012, Harris_2007, Jarzynski_2011, Campisi_2011, Esposito_2009, Manzano_2022, strasberg2022quantum, Ciliberto_2017} aims to understand the thermodynamic behavior of nanoscale systems out of equilibrium. In this framework, thermodynamic quantities, such as entropy, work, and heat are fluctuating quantities defined at the trajectory level. A crucial result in this field is provided by the fluctuation theorem (FT), $\langle e^{-\sigma} \rangle = 1$~\cite{Jarzynski_1997a, Jarzynski_1997b, Jarzynski_1999, Crooks_1998, Crooks_1999, Crooks_2000, Seifert_2005, Tasaki_2020, Kurchan_2000, Bochkov_1977, BOCHKOV1, BOCHKOV2, Manzano_2018, TalknerLutz, rodrigues2023nonequilibrium}, where  $\sigma$ denotes the entropy production and $\langle \cdot \rangle$ the ensemble average. This relation implies the second law of thermodynamics, $\langle \sigma \rangle \geq 0$, and therefore, it can be considered a generalization thereof.

%{ \color{blue} Crucially, the above results hold both in the \textit{classical} and \textit{quantum} regime, which are very different. In the \textit{classical} one, applicable to incoherent systems, the system follows a phase-space trajectory~\cite{Jarzynski_1997a, Seifert_2005} In contrast, in the \textit{quantum} regime that applies to systems with quantum coherence, we have to define the stochastic entropy production along a quantum jump or two-point measurement trajectory~\cite{Manzano_2018, Manzano_2022}}

%{ \color{blue} Moreover, when generalizing the FT to systems with measurement and feedback, which is a topic of this paper, the measurement in the \textit{quantum} regime causes a backaction on the state of the system, which is absent in the \textit{classical} regime, making them so distinct.}
The idea of employing measurement and feedback in thermodynamic processes dates back to the thought experiments of Maxwell and Szilard~\cite{maxwell1872theory, Szilard1929berDE, Rex}, where knowledge about the microscopic degrees of freedom could be utilized to seemingly overcome the second law of thermodynamics. The FT and the second law can be generalized to feedback controlled processes by treating information on an equal footing as entropy production~\cite{ParrondoHorowitzSagawa}, resulting in FTs that include a stochastic information term, $I$, resulting in~\cite{Sagawa_2008, Sagawa_2010, Sagawa_2012, Sagawa_2012b, Horowitz_2010, Ito_2013, Potts_2018, Ponmurugan_2010, Morikuni_2010, Funo_2013, Gong_2016, Murashita_2017, Ashida_2014, Funo_2015, Ashida_2014, Yada_2022, Sagawa_2013b, Abreu_2012, Lahiri_2012, Wachtler_2016, Shiriashi_2015, Schmitt, HorowitzEsposito_2014, Shiraishi_2015b, Tobaye, Koski, archambault2024fp, Naghiloo, Masayuma, Camati, Paneru}
\begin{equation}
\label{eq:FT intro}
    \langle e^{-\sigma -I } \rangle = 1 \implies     \langle \sigma  \rangle  \geq - \langle I  \rangle,
\end{equation}
which allows $\langle \sigma \rangle$ to become negative. Crucially, for a given feedback control protocol there is no unique way to define the information term $I$~\cite{Potts_2018}. %since it is only bounded by the negative average of the information term.
FTs including information are of particular interest because, on the one hand, they shed insight into the interplay between thermodynamics and information. On the other hand, they provide constraints on monitored nanoscale systems that are promising for emerging nano and quantum technologies. 

Remarkably, FTs continue to hold in the quantum regime, even though there are no well-defined trajectories through phase space. Instead, trajectories can be defined using quantum jumps or projective measurements of the initial and final states~\cite{Manzano_2018, Manzano_2022}. This conceptual difference between the classical and the quantum regime is exacerbated in the presence of measurements and feedback, due to the crucial role of measurement backaction in the presence of quantum coherence. For these reasons, a FT for arbitrary measurement and feedback schemes in the quantum regime is still lacking.

In this Letter, we fill this gap by deriving an integral FT that holds for open quantum systems undergoing arbitrary measurement and feedback,
\begin{equation}
\label{eq: FT main}
        \langle e^{-(\sigma - \sigma_{\rm cg}) } \rangle = 1,
\end{equation}
as well as a generalized second law,
\begin{equation}
\label{eq: SL main}
    \langle \sigma \rangle \geq  \langle \sigma_{\rm cg} \rangle,
\end{equation}
where the information term, $I = -\sigma_{\rm cg}$, is given by the so-called coarse-grained entropy~\cite{Kawai_2007}, an experimentally accessible quantity that depends only on the measurement outcomes $Y$. To obtain Eqs.~\eqref{eq: FT main} and~\eqref{eq: SL main} we derive a detailed FT below.
%{\color{blue} Remarkably, Eqs.~\eqref{eq: FT main} and \eqref{eq: SL main} have the same form as classical FT and SL with coarse-grained entropy in Ref.~\cite{Potts_2018}.
 %This means that the coarse-grained entropy, which can be understood as the entropy production inferable from the measurement outcomes only, naturally emerges as an information term even in the presence of measurement backaction, which is not taken into account in the classical regime. Extending these results from the \textit{classical} to the \textit{quantum} regime is not trivial as a result of the conceptual differences between them.}

\begin{figure*}[t]
    \centering
    \includegraphics[width=0.8\textwidth]{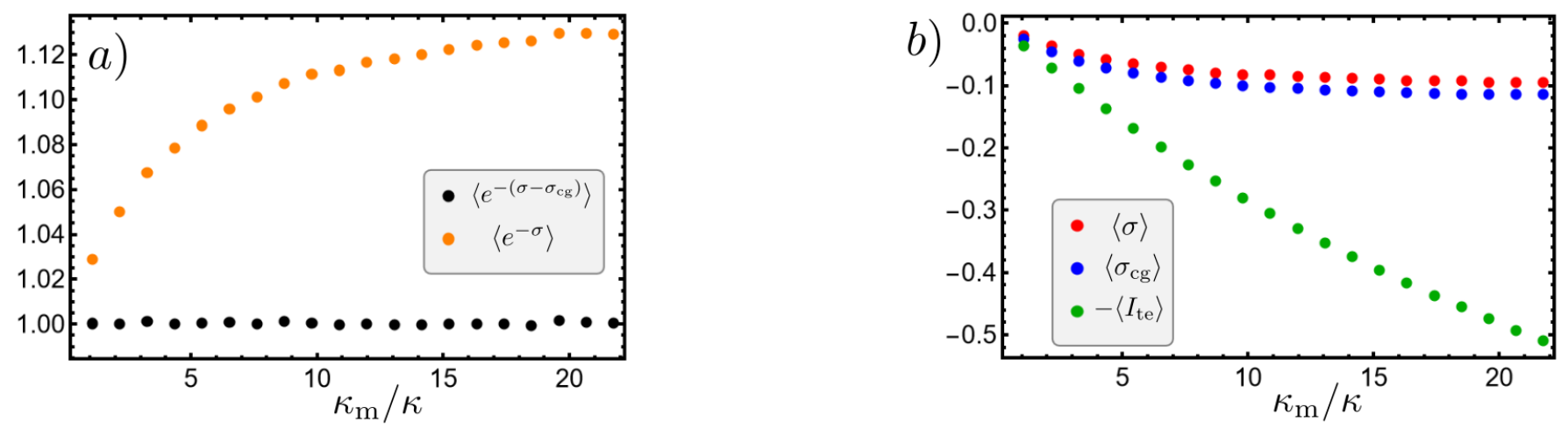}
    \caption{Fluctuation theorem and second law for continuous feedback control. (a) Conventional FT $\langle e^{-\sigma} \rangle $ (orange) and our FT $\langle e^{-(\sigma - \sigma_{\rm cg})} \rangle $ (black) as a function of the measurement rate $\kappa_{\rm m}$. While the conventional FT breaks down, Eq.~\eqref{eq: FT main} holds.  (b) The entropy production $\langle \sigma \rangle$ (red) is compared to the bounds provided by the coarse-grained entropy $\langle \sigma_{\rm cg} \rangle$ (blue), and the quantum-classical transfer entropy $-\langle I_{\rm te} \rangle$ (green). The coarse-grained entropy provides a useful bound for all measurement rates, while the quantum-classical transfer entropy becomes uninformative in the limit of large measurement rates. Parameters: $\tau/\beta = 10$, $\beta \omega = 0.3$,  $\beta \chi = 0.04$, $\beta\epsilon = 0.2$, $\beta \Omega = 0.1 \pi$, $\kappa/\omega = 0.116$. Each data point was obtained using a quantum-jump simulation with $10^5$ trajectories.}
    \label{fig:ContinuousToyModel}
\end{figure*}

\emph{Comparison with previous results--}
Our work substantially goes beyond previous FTs including measurement and feedback in the quantum regime, which mainly focused on single feedback \cite{Morikuni_2010,Funo_2013,Funo_2015,Gong_2016,Murashita_2017}. A notable exception is Ref.~\cite{Yada_2022}, which derived a FT for continuously observing quantum jumps. While constituting an important step toward a FT for arbitrary measurement and feedback schemes, the corresponding FT has significant shortcomings, limiting its applicability: (i) the information term, provided by the quantum-classical transfer entropy ($I_{\rm te}$), a generalization of the mutual information, cannot be experimentally sampled without the aid of numerical simulations, since it depends explicitly on the system state along a given trajectory. We note that this shortcoming also applies to Refs.~\cite{Funo_2013,Funo_2015,Gong_2016,Murashita_2017} (ii) The FT is restricted to Markovian systems under quantum-jump monitoring, which excludes non-Markovian scenarios and other continuous measurements such as charge sensing. (iii) In the case of strong and continuous measurements, $\langle I_{\rm te} \rangle$ can significantly outgrow $\langle \sigma \rangle$, causing the corresponding second law to become inapplicable [see Fig.~\ref{fig:ContinuousToyModel}~(b)]. This is analogous to classical systems, where the transfer entropy~\cite{Sagawa_2012} may also diverge~\cite{Horowitz_2014, Schmitt}.

%A lack of FT and a second law valid for quantum systems under arbitrary continuous and discreet measurement and feedback constitutes a major gap in the understanding of thermodynamics of nonequilibrium systems. In a recent paper~\cite{Yada_2022}, Yada and co-authors made a step to address this problem and derived a FT and a second law for Markovian open quantum systems with continuous quantum-jump measurements. The corresponding information term $I$ is given by the quantum-classical transfer entropy (here denoted as $I_{\rm te}$), which generalizes the transfer entropy for classical systems (see Refs.~\cite{Sagawa_2012, Horowitz_2010}).

Our results, which may be applied to discrete and continuous feedback, circumvent the above-mentioned shortcomings: (i) the coarse-grained entropy depends only on the measurement outcomes and can therefore be experimentally sampled. (ii) Our findings are derived using the most general descriptions for measurements (POVMs) and system dynamics (unitary evolution of system and environment), making them applicable to both Markovian and non-Markovian scenarios. (iii) The coarse-grained entropy does not diverge (except for diverging entropy production, when FTs break down in general). As such, $\langle \sigma_{\rm cg} \rangle$ provides a useful bound on $\langle \sigma \rangle$, even when $\langle I_{\rm te} \rangle$ diverges. To illustrate this, we numerically investigate a continuous measurement of a qubit [see Fig.~\ref{fig:ContinuousToyModel} ~(b)]. 

Finally, we note that deriving our quantum FT with coarse-grained entropy does not require any fictitious measurements before and after measurements, which are necessary to extend the mutual information to the quantum-classical transfer entropy for quantum systems~\cite{SM} as a result of the measurement backaction. %{\color{blue} A more detailed explanation of the quantum-classical transfer entropy is included in the SM.}

Our work thus provides an experimentally testable FT and second law, valid for open quantum systems under arbitrary continuous and discrete measurement and feedback schemes.

\emph{General measurement and feedback scheme--}
In contrast to classical systems, where FTs in the presence of measurement and feedback can be derived using conditional probabilities and Bayes theorem \cite{Sagawa_2012,Potts_2018}, the invasive role of measurements in quantum theory requires us to consider a more explicit model. Remarkably, we nevertheless find a detailed FT that has the same form as in the classical regime \cite{Potts_2018}.

We consider an open quantum system that may exchange energy with a reservoir at an inverse temperature $\beta$. At times $t_n$, the system is being measured and the outcome $y_n$ is obtained. The system and reservoir together are described by a density matrix $\hat{\varrho}^{Y_n}_t$, conditioned on all previous measurement outcomes $Y_n \equiv \{y_1, y_2, ... , y_n \}$, with $t_n<t<t_{n+1}$. The measurements are modeled by POVMs~\cite{nielsen2010quantum, wiseman2010quantum, jacobs_2006} and update the state according to 
\begin{equation}
    \label{eq:povm}
    \hat{\varrho}^{Y_n}_{t_n} = \frac{\hat{M}_{n}(y_n) \hat{\varrho}^{Y_{n-1}}_{t_n} \hat{M}^{\dagger}_{n}(y_n)}{ P[y_n|Y_{n-1}]},
\end{equation}
where $P[y_n|Y_{n-1}] = \text{Tr} \{ \hat{M}^{\dagger}_{n}(y_n) \hat{M}_{n}(y_n) \hat{\varrho}^{Y_{n-1}}_{t_n}  \}$ and $\hat{M}_{n}(y_n)$ is the Kraus operator corresponding to obtaining outcome $y_n$ in the measurement at $t_n$. The set of Kraus operators fulfils the relation $\sum_{y} \hat{M}^{\dagger}_n(y) \hat{M}_n(y) = \hat{I} $, where $\hat{I}$ is the identity operator.

In between measurements, the time evolution of the system and reservoir together is unitary and determined by the total Hamiltonian $\hat{H}_{\rm tot}(\lambda_t^{Y_n}) = \hat{H}(\lambda_t^{Y_n})+\hat{V}(\lambda_t^{Y_n})+\hat{H}_{\rm R}$, where the first term denotes the Hamiltonian of the system, the second term the system-reservoir coupling, and the last term the Hamiltonian of the reservoir. The Hamiltonian depends on a control parameter $\lambda_t^{Y_n}$, which may depend on both time and the set of previous measurement outcomes. %For convenience, with $\Lambda_{Y_n}$ we denote $\lambda_t^{Y_n}$ in the entire duration of the protocol for a given $Y$. A quantum state of the system at a particular time is given by $\op{\rho}_t^Y = \text{Tr}_{\rm R}\{\op{\varrho}_{t}^Y \}$.

At the initial time $t = 0$, the joint quantum state is given by $\op{\varrho}_{0} = \op{\rho}_0 \otimes \op{\tau}_{\rm R}$, where $\op{\rho}_0= \sum_a p_a \proj{a}$ is the initial state of the system, and $\op{\tau}_{\rm R} = e^{-\beta \op{H}_{\rm R}}/Z_{\rm R} $ is the thermal state of the reservoir with the inverse temperature $\beta$ and the partition function $Z_{\rm R} = \text{Tr} \{ e^{-\beta \op{H}_{\rm R}} \}$. At all times, the reduced state of the system may be obtained by tracing out the reservoir $\op{\rho}_t^{Y_n} = \text{Tr}_{\rm R}\{\op{\varrho}_{t}^{Y_n} \}$. For simplicity, we assume that at the final time $t= \tau$ the control parameter takes the same value $\lambda_\tau$ for all $Y \equiv Y_M$, where $M$ denotes the total number of measurements. However, our approach can readily be generalized (see~\cite{SM}).

\emph{Trajectory thermodynamics--}
The evolution of the quantum state of the system and the reservoir can be unraveled into the trajectories $\Gamma= \{Y, \gamma\}$ containing the set of the measurement outcomes $Y$ and information about the system and the reservoir $\gamma = \{a, E_0, f, E_\tau\}$ at the beginning and the end of the protocol. Here, $E_0$ and $E_\tau$ are eigenvalues of $\hat{H}_{\rm R}$ and refer to the energies of the reservoir at the beginning and at the end of the trajectory. Similarly, $a$ and $f$ denote the system state at the beginning and end of the trajectory. While $a$ is an eigenvalue of the initial state $\op{\rho}_0$, $f$ is an eigenvalue of the "reference state" $\op{\rho}_{\rm r} = \sum_f p_f \proj{f}$, which in general may be chosen arbitrarily. As we will see below, it specifies the initial quantum state of the system in the backward experiment. Here, we restrict ourselves to the case where $\op{\rho}_{\rm r}$ is independent of $Y$ (see~\cite{SM} for a generalization).

The entropy production corresponding to the trajectory $\Gamma$ is defined as~\cite{Manzano_2018, Manzano_2015, Funo_2015, Landi_2021, Campisi_2011}
\begin{equation}
\label{eq: sigma def}
        \sigma[\Gamma] = -\ln{p_f} + \ln{p_a} + \beta Q[\Gamma],
\end{equation}
where $ Q[\Gamma]  = E_\tau - E_0$ is the change of the energy of the reservoir, which we identify as heat. %The ensemble average is given by $\langle \sigma \rangle = \sum_\Gamma P[\Gamma] \sigma[\Gamma]$, where $P[\Gamma]$ is the probability of $\Gamma$. The average energy change of the reservoir reads $\langle Q \rangle = \sum_{Y} P[Y] \text{Tr} \{ \op{H}^R \op{\varrho}_\tau^Y \} - \text{Tr} \{\op{H}^R \op{\varrho}_{0} \}$, where the probability of obtaining the set of the measurement outcomes $Y$ is given marginal distribution $P[Y] = \sum_{\gamma} P[\Gamma]$.
We denote the probability associated to a trajectory by $P[\Gamma]$. It can be computed by choosing $|a\rangle\otimes |E_0\rangle$ as the initial state, computing the total density matrix conditioned on the measurement outcomes $Y$, and projecting onto the state $|f\rangle\otimes |E_\tau\rangle$~\cite{SM}.

For the reference state, two choices are popular. First, one may choose the average final state $\op{\rho}_\mathrm{r} = \sum_{Y} P[Y] \op{\rho}_\tau^{Y}$, such that the average of $\sigma[\Gamma]$ reduces to the average entropy production in the measurement and feedback experiment; see, for instance,~\cite{Yada_2022, Funo_2015}. Second, one may choose a thermal state $\op{\rho}_{\rm r} = e^{-\beta \hat{H}(\lambda_\tau)}/\text{Tr} \{ e^{-\beta \hat{H}(\lambda_\tau)} \}$, such that, when the initial state is also thermal, the FT results in a generalization of the Jarzynski relation; see, for instance,~\cite{Gong_2016, Funo_2015}.
%The stochastic entropy production and its ensemble average is specified with the reference state. The first choice we consider is when $\op{\rho}^r = \op{\rho}_\tau$, where $ \op{\rho}_\tau :=  \sum_{Y} P[Y] \op{\rho}_\tau^{Y}$ is the average final state of the system. The corresponding entropy production reads
%\begin{equation}
%\label{eq: sigma S Q}
%    \langle \sigma \rangle = S(\op{\rho}_\tau) - S(\op{\rho}_\tau) + \beta \langle Q \rangle,
%\end{equation}
%where $S(\op{\rho}) = -\text{Tr} \{ \op{\rho} \ln{\op{\rho}} \}$ is the von Neumann entropy of the system.
%Alternatively, we can prepare the initial and reference states as the thermal equilibrium states, $\op{\rho}_0 = e^{-\beta \hat{H(\lambda_0)}}/Z_0$ and $\op{\rho}_{\rm r} = e^{-\beta \hat{H(\lambda_\tau)}}/Z_\tau$, where $Z_t = \text{Tr} \{ e^{-\beta \hat{H(\lambda_t^Y)}} \}$. The corresponding average entropy production is given by the dissipated work
%\begin{equation}
%\label{eq: sigma diss work}
%    \langle \sigma \rangle = \beta \left( U_\tau -  F(\lambda_\tau) - U_0  +   F(\lambda_0)  +  \langle Q \rangle \right),
%\end{equation}
%where $F(\lambda_t) = - (\ln{Z_t})/\beta$ is the equilibrium free energy of the system, and $U_0= \text{Tr}\{ \op{H}(\lambda_0) \op{\rho}_0 \}$ and $U_\tau = \sum_Y P[Y] \text{Tr}\{ \op{H}(\lambda_\tau) \op{\rho}_t^Y \}$ are the average initial and final energies of the system, respectively.

\begin{figure*}[t]
    \centering
    \includegraphics[width=0.98\textwidth]{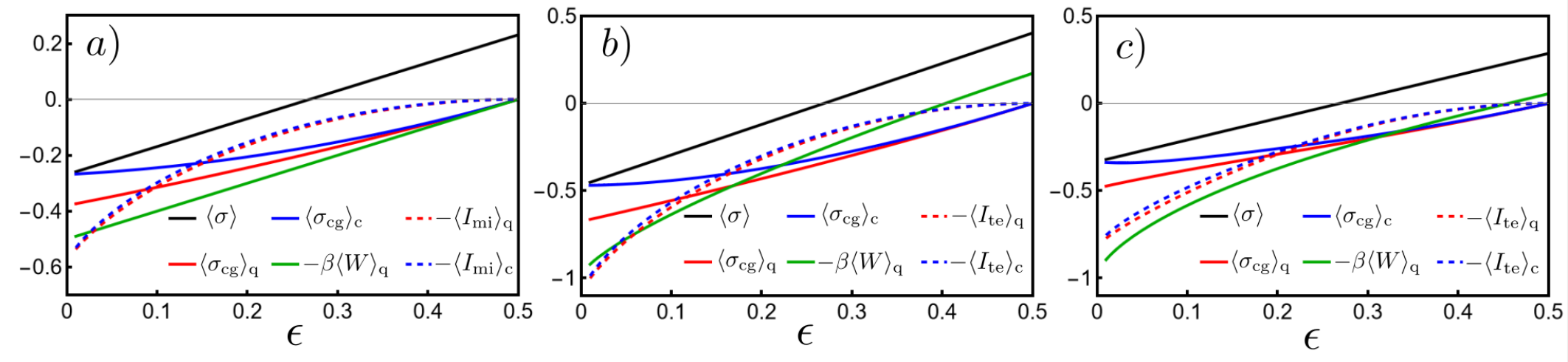}
    \caption{The second law for the qubit model. Entropy production $\langle \sigma \rangle$ (black), coarse-grained $\langle \sigma_{\rm cg} \rangle_{\rm c/q}$ (blue, red), (quantum-classical) mutual information [panel (a)] or transfer entropy [panels (b) and (c)] $\langle I_{\rm mi/te} \rangle_{\rm c/q}$ (blue-dashed, red-dashed), and extracted work $ \langle W \rangle_{\rm q}$ (green) as a function of the measurement error $\epsilon$. The subscripts "c" and "q" correspond to the classical and quantum protocols, respectively. The entropy production $\langle \sigma \rangle$ is equal for both cases but only for the classical protocol can it be related to the extracted work $\beta \langle W \rangle_{\rm c}  = -\langle \sigma \rangle$ . (a) Single measurement. (b) Two measurements with $\kappa \Delta t =1$. (c) Two measurements with $\kappa \Delta t =0.2$. In all panels, $\beta \omega = 1$.}
    \label{fig:QubitToyModel}
\end{figure*}

\emph{FTs and the second law.}
In order to derive the detailed FT [see Eq.~\eqref{eq: detailed FT main}], we need to introduce the backward experiment. In the backward experiment, measurements with corresponding Kraus operators $\hat{\Theta} \op{M}^\dagger_n(y) \hat{\Theta}^{-1}$, where $\hat{\Theta}$ is the time-reversal operator~\cite{Campisi_2011}, are performed at times $\tau-t_n$. Note that these Kraus operators always define elements of POVMs and thus correspond to physical measurements~\cite{SM}. In between measurements, the time evolution is determined by the Hamiltonian $\hat{\Theta}\hat{H}_{\rm tot}(\xi_t)\hat{\Theta}^{-1}$. In contrast to the forward experiment, no feedback is performed in the backward experiment. Instead, a fixed protocol $\{\xi_t\}$ is applied. The initial state for the backward experiment is given by the time-reversed reference state $\hat{\Theta} \op{\rho}_{\rm r}  \otimes \op{\tau}_{\rm R} \hat{\Theta}^{-1}$. We denote the probability for a trajectory $\Gamma$ given a fixed protocol as $P_{\rm tr}[\Gamma|\{\xi_t\}]$, where the subscript stands for "time-reversed". Before introducing the full backward experiment, we note that the following FT holds:
\begin{equation}
\label{eq: FT sigma}
    e^{-\sigma[\Gamma]} = \frac{P_{\rm tr}[\bar{\Gamma}|\{{\lambda}_{\tau-t}^Y\}]}{P[\Gamma]},
\end{equation}
where the overbar signifies time reversal, i.e., $\bar{\Gamma} = \{\bar{Y}, \bar{\gamma} \}$, with $ \bar{\gamma} =  \{f, E_\tau, a, E_0 \}$ and $\bar{Y} = \{y_M, ... , y_1\}$. This is in fact the standard FT~\cite{Manzano_2018, Manzano_2015, Campisi_2011, Funo_2015}, since $P[\Gamma]$ also provides the probability for the trajectory $\Gamma$ in an experiment where the protocol $\{{\lambda}_{t}^Y\}$ is determined in advance rather than by feedback. In addition, it is known that adding measurements (without feedback) does not alter the FT~\cite{Wanatabe_2014, Campisi_2010, Campisi_2011b}

%For a given trajectory $\Gamma$ in the forward experiment, we define its reversal as $\bar{\Gamma} = \{\bar{Y}, \bar{\gamma} \}$, where $ \bar{\gamma} =  \{f, E_\tau, a, E_0 \} \}$ and $\bar{Y} = \{\bar{y}_1 = y_M, ... , \bar{y}_M = y_1\}$ denotes the set of POVM measurement outcomes that correspond to the Kraus operators $\op{M}_{\bar{y}_i} := \Theta \op{M}_{y_{M+1-i}}^\dagger \Theta^\dagger$.  

%We define the probability $P[\bar{\Gamma} | \bar{\Lambda}_Y ]$ that the system follows the inverse trajectory $\bar{\Gamma}$ if we implement the time reversal of the protocol $\Lambda(Y)$, which we denote with $\bar{\Lambda}_Y$. The inverse protocol $\bar{\Lambda}_Y$ is realised by preparing the quantum state $\Theta \op{\rho}_{\rm r} \Theta^{\dagger} \otimes \Theta \op{\tau}_B \Theta^{\dagger}$ and allowing it to evolve in time according to the equation
%\begin{equation}
%\label{eq: von Neumann backward} 
%\frac{d}{dt} \op{\rho}_{t, tot}^Y = -i[\bar{H}(\bar{\lambda}_t^Y) + \bar{H}_I(\bar{\lambda}_t^Y) + \bar{H}_B , \op{\rho}_{t, tot}^Y ],
%\end{equation}
%where $\bar{\lambda}_t^Y = \lambda_{\tau - t}^Y$, $\bar{O} = \Theta \hat{O} \Theta^{\dagger}$, and $\Theta$ is the quantum time-reversal operator. At the times $\{\tau - t_{m_M}, ..., \tau - t_{m_{M+1 - i}}, ..., \tau - t_{m_1}\}$ we make the POVM measurements, such that for the $i$-th measurement the set of Kraus operators that realises it includes $\op{M}_{\bar{y}_i}$. Importantly, we do not perform any feedback in the inverse protocol.

Similar to the classical case in Ref.~\cite{Potts_2018}, the full backward experiment is then described by the distribution
\begin{equation}
\label{eq: PB}
    P_{\rm B}[\bar{\Gamma}] = \frac{ P_{\rm tr}[\bar{\Gamma}|\{{\lambda}_{\tau-t}^Y\}] P[Y] }{P_{\rm tr}[\bar{Y} | \{{\lambda}_{\tau-t}^Y\} ]},
\end{equation}
where $ P[Y]=\sum_\gamma P[\Gamma]$ is the probability that the measurement outcomes $Y$ are obtained in the forward experiment and $P_{\rm tr}[\bar{Y} | \{{\lambda}_{\tau-t}^Y\} ]=\sum_\gamma P_{\rm tr}[\bar{\Gamma}|\{{\lambda}_{\tau-t}^Y\}]$ is the probability of obtaining $\bar{Y}$ in the time-reversed scenario. The backward experiment determined by Eq.~\eqref{eq: PB} has a clear experimental interpretation: (1) $Y$ is sampled from the distribution $P[Y]$. (2) The protocol $\{{\lambda}_{\tau-t}^Y\}$ is implemented, together with the time-reversed measurements. (3) A postselection is performed: if the measurement outcomes coincide with $\bar{Y}$, the experiment is a success; otherwise, the data is discarded and the experiment repeated starting from step 2.

To derive a detailed FT, we note that (see~\cite{SM} for a derivation and definition)
\begin{equation}
\label{eq: sigma cg m}
    e^{-\sigma_{\rm cg}[Y]} = \frac{P_{\rm tr}[\bar{Y} | \{{\lambda}_{\tau-t}^Y\} ]}{P[Y]},
\end{equation}
where $\sigma_{\rm cg}[Y]$ denotes the entropy production, coarse-grained over $\gamma$, the experimentally inaccessible part of the trajectory~\cite{Sagawa_2012, Kawai_2007, Potts_2018}. By comparing the last equation with Eq.~\eqref{eq: FT sigma}, we may interpret $\sigma_{\rm cg}[Y]$ as the entropy production inferable from the measurement outcomes alone. 
From Eqs.~(\ref{eq: FT sigma})-(\ref{eq: sigma cg m}), we find a detailed fluctuation theorem, our first main result,
\begin{equation}
\label{eq: detailed FT main}
    \frac{P_{\rm B}[\bar{\Gamma}]}{P[\Gamma]}  = e^{-(\sigma[\Gamma] - \sigma_{\rm cg}[Y])}.
\end{equation}
As mentioned above, this equation has the same form as the corresponding FT in the classical regime~ \cite{Potts_2018}.
Using Jensen's inequality $\langle f(X)\rangle \leq f(\langle X \rangle)$ for a convex function $f(X)$, we obtain the second law of information thermodynamics given in Eq.~\eqref{eq: SL main}, which constitutes our second main result.
%These findings demonstrate that the coarse-grained entropy plays a role of the information term in the FT and the second law for quantum systems under measurement and feedback. This quantity does not diverge as long as $\sigma[\Gamma]$ remains finite, which addresses the shortcoming i). Moreover, both $P[Y]$ and $P[\bar{Y}|\bar{\Lambda}_Y]$ have a clear operational meaning and directly depend on the measurement outcomes, but not on the system. This makes the coarse-grained entropy accessible for each trajectory from the measurement apparatus, thus solving the shortcoming ii). 

%The coarse-grained entropy is defined as~\cite{Sagawa_2012, Kawai_2007, Potts_2018}
%\begin{equation}
%\label{eq: sigma cg def}
%    e^{-\sigma_{\rm cg}[Y]} :=  \sum_{\gamma} e^{-\sigma[\Gamma]} P[\gamma | Y],
%\end{equation}
%where $P[\gamma | Y] = P[\Gamma]/P[Y]$ is the conditional probability. Using our definition of the backward experiment, we can express it as
%\begin{equation}
%\label{eq: sigma cg m}
%    e^{-\sigma_{\rm cg}[Y]} = \frac{P[\bar{Y}|\bar{\Lambda}_Y]}{P[Y]}.
%\end{equation}
%By noticing that the entropy production in Eq.~\eqref{eq: sigma def} can be written as $e^{-\sigma[\Gamma]} = P[\bar{\Gamma}| \bar{\Lambda}_Y ]/P[\Gamma]$ and using Eq.~\eqref{eq: sigma cg m}, we obtain the detailed FT in Eq.~\eqref{eq: detailed FT main}, as well as the integral FT in Eq.~\eqref{eq: FT main}. These two equations are our first main result of this Letter. 

\emph{Lindblad master equation--}
Since open quantum systems are very often modeled using Lindblad master equations, we now adapt the results of the previous section to scenarios where the dynamics of the reduced density matrix in between the measurements ($t_n < t < t_{n+1}$) is described by~\cite{Breuer, Potts_2021, Albash_2012, landi_2023} (we set $\hbar =1$ hereafter)
\begin{equation}
\label{eq:Lindblad forward}
    \frac{d}{dt} \op{\rho}_t^{Y_n} =   -i[\hat{H}(\lambda_t^{Y_n}) , \op{\rho}_t^{Y_n} ] + \sum_j \mathcal{D}[\op{L}_j(\lambda_t^{Y_n})]\op{\rho}_t^{Y_n} 
\end{equation}
where $\mathcal{D}[\hat{O}]\op{\rho} = \hat{O} \op{\rho} \hat{O}^{\dagger} - \frac{1}{2} \{  \op{\rho}, \hat{O}^{\dagger}\hat{O}  \} $. The POVM measurements update the density matrix $\op{\rho}_{t_n}^{Y_n}$ according to the rule presented in Eq.~\eqref{eq:povm}. The jump operators %are ladder operators of a thermodynamic Hamiltonian $\op{H}_{\rm TD}$ and fulfil the relation $[\op{L}_j (\lambda_t^Y) , \op{H}_{\rm TD} (\lambda_t^Y) ] = Q_j (\lambda_t^Y) \op{L}_j (\lambda_t^Y)$, where $ Q_j(\lambda_t^Y)$ is the heat dissipated to the reservoir associated with the $j$-th jump~\cite{Potts_2021}. They 
obey a local detailed balance, i.e., for each $j$ there is a $\tilde{j}$ such that $\op{L}_{\tilde{j}} = \op{L}_j^{\dagger} e^{-\beta q_j/2}$, with $q_j = -  q_{\tilde{j}}$.  

%When the dynamics of the system is well described with the Lindblad master equation, all our results can be equivalently derived without explicitly incorporating the reservoir. To this end, the evolution of the system during the feedback protocol can be unravelled into the stochastic quantum jump trajectories that correspond to the different paths the system can take~\cite{Manzano_2022, Manzano_2018, Yada_2022, Horowitz_2012, Horowitz_2013, Gong_2016, wiseman2010quantum}. 
In this scenario, a trajectory is defined as $\Gamma = \{Y, \gamma\}$, where $\gamma = \{a, (s_1, j_1),..., (s_k, j_k),..., (s_K, j_K), f \}$. Here $(s_k,j_k)$ implies that at time $s_k$, a jump mediated by the operator $\op{L}_{j_k}$ occurred. The remaining symbols retain their meaning. On the trajectory level, the heat released to the reservoir reads $Q[\Gamma] = \sum_k  q_{j_k}(\lambda_{s_k}^Y) $. The expression for $\sigma[\Gamma]$ given in Eq.~\eqref{eq: sigma def}, as well as the associated ensemble averages continue to hold~\cite{Horowitz_2012, Horowitz_2013, Gong_2016, Manzano_2018, Yada_2022}.% in the stochastic quantum jump unravelling of the Lindblad master equation. 

The backward experiment and the probability $P_{\rm tr}[\Gamma|\{ \xi_t\}]$ are analogous to the unitary case (see~\cite{SM} for details). In the quantum-jump unraveling, the reversed version of $\Gamma$ is given by $\bar{\Gamma} = \{\bar{Y}, \bar{\gamma} \}$, where $ \bar{\gamma} =  \{f, (\tau - s_K, \tilde{j}_K),..., ( \tau - s_{k},  \tilde{j}_{k}) ,..., (\tau - s_1, \tilde{j}_1), a \}$, which results in Eq.~\eqref{eq: FT sigma}. By defining the full backward experiment with Eq.~\eqref{eq: PB}, we recover our main results, Eqs.~\eqref{eq: detailed FT main}, \eqref{eq: FT main}, and \eqref{eq: SL main} for systems described by Lindblad master equations. %FTs and the second law of thermodynamics with the coarse-grained entropy, which similarly can be expressed using Eq.~\eqref{eq: sigma cg m}. 

%\begin{equation}
%\label{eq:Lindblad backward}
%    \frac{d}{dt} \op{\rho}_t =  -i[\bar{H}(\bar{\lambda}_t^Y), \op{\rho}(t) ] + \sum_k \mathcal{D}[\bar{L}_k(\bar{\lambda}_t^Y)]\op{\rho}_t,
%\end{equation}

\emph{Qubit model--}
We illustrate our results by investigating a feedback controlled qubit with a Hamiltonian $\op{H} = \frac{\omega}{2} \op{\sigma}_z$, where $\op{\sigma}_{x, y , z}$ are Pauli matrices in the basis $\ket{0}$ (ground) and $\ket{1}$ (excited). The system is initially in a thermal state with inverse temperature $\beta$, $\hat{\rho}_0 = e^{-\beta \op{H}}/\text{Tr}\{ e^{-\beta \op{H}} \}$. We consider multiple feedback protocols that aim at extracting work from the qubit. First, we consider a single measurement with Kraus operators that commute with the Hamiltonian: $ \op{M}_0 = \sqrt{1-\epsilon} \proj{0} + \sqrt{\epsilon} \proj{1} $ and $ \op{M}_1 = \sqrt{1-\epsilon} \proj{1} + \sqrt{\epsilon} \proj{0}$. If the measurement outcome is $0$, the system is assumed to be in the ground state and no feedback is performed. If the measurement outcome is equal to $1$, the system is assumed to be in the excited state and the unitary transformation $\op{U}_1 = \ket{0} \bra{1} + \ket{1} \bra{0}$ is applied. This transformation can extract work by transforming the excited state to the ground state. %Should a measurement error have occured, which happens with probability $\epsilon$, the transformation instead performs work on the system. After the feedback, the system thermalizes again, making the protocol cyclic. 
In this protocol, heat from the bath is turned into work using information. Since only the populations of the qubit are relevant, we label this protocol as classical. Considering as a reference state the thermal state, the first law of thermodynamics implies that the entropy production is determined by the extracted work $\langle \sigma\rangle = -\beta \langle W \rangle_{\rm c}$, where the subscript stands for classical. Equation \eqref{eq: SL main} then provides a bound on the extracted work.
This is illustrated in Fig.~\ref{fig:QubitToyModel}~(a), where we compare our bound to the mutual information $\langle I_{\rm mi}\rangle_{\rm c}$~\cite{Sagawa_2010}. Notably, for small measurement errors, our bound is tighter.
%This is illustrated in Fig.~\ref{fig:QubitToyModel}\,a), where we compare our bound to the mutual information $\langle I_{\rm mi}\rangle_{\rm c}$. 

%Before applying a feedback protocol, which consists of a POVM measurement and a corresponding unitary transformation, the system is isolated from the reservoir. We consider two types of the measurement: classical, where the Kraus operators commute with the Hamiltonian, and quantum. In the classical case, the operators are given by $ \op{M}_0 = \sqrt{1-\epsilon} \proj{0} + \sqrt{\epsilon} \proj{1} $ and $ \op{M}_1 = \sqrt{1-\epsilon} \proj{1} + \sqrt{\epsilon} \proj{0}$, whereas the corresponding unitary transformations read $U_0 = \op{I}$ and $U_1 = \ket{0} \bra{1} + \ket{1} \bra{0}$. 

In addition to the classical protocol, we consider a quantum protocol where the measurement operators do not commute with the Hamiltonian: $ \op{M}_+ = \sqrt{1-\epsilon} \proj{+} + \sqrt{\epsilon} \proj{-} $ and $ \op{M}_- = \sqrt{1-\epsilon} \proj{-} + \sqrt{\epsilon} \proj{+}$, where $\ket{\pm} = (\ket{0} + \ket{1})/\sqrt{2}$. The corresponding unitary transformations employed to extract work are $\op{U}_+ = \ket{0} \bra{+} + \ket{1} \bra{-}$ and $\op{U}_- = \ket{0} \bra{-} + \ket{1} \bra{+}$. In this protocol, the measurement provides the energy that is then extracted in the form of work, in analogy to Refs.~\cite{Elouard_2017_engine, Elouard_2017_qh}. This implies that the extracted work can no longer be related to the heat exchanged with the bath and generally differs from the entropy production. As illustrated in Fig.~\ref{fig:QubitToyModel}~(a), more work can be extracted with the measurement providing an additional source of energy. Because of the similarity of the classical and the quantum protocol, we find that the entropy production is equal in the two cases. However, in the quantum protocol our bound becomes slightly less tight because the measurement outcome provides less information on heat, which fully determines the entropy production.
%{\color{blue} The bound provided by the quantum-classical mutual information $\langle I_{\rm mi}\rangle_{\rm q}$, where the subscript stands for quantum protocol, hardly changes compared to the classical protocol.}
The bound provided by the quantum-classical mutual information $\langle I_{\rm mi}\rangle_{\rm q}$, which generalizes the mutual information to the quantum regime~\cite{Sagawa_2008}, hardly changes compared to the classical protocol.

Similar observations can be made when measurement and feedback is applied twice, with a time delay $\Delta t$ in between the measurements. During this intermediate stage the interaction with the reservoir is described with the Lindblad jump operators $\op{L}_{-} = \sqrt{\kappa} \op{\sigma}_- $ and $\op{L}_{+} = \sqrt{\kappa} e^{-\beta\omega/2} \op{\sigma}_+ $, where $\op{\sigma}_\pm=(\op{\sigma}_x\pm i \op{\sigma}_y)/2$.
The second law is shown in Figs.~\ref{fig:QubitToyModel}~b) and~(c) for $\kappa \Delta t = 1$ and $\kappa \Delta t = 0.2$, respectively. %{\color{blue} DELETED SENTENCE}
%In this case, $\langle I_{\rm te} \rangle_{\rm c}$ denotes the transfer entropy, and $\langle I_{\rm te} \rangle_{\rm q}$ denotes the quantum-classical transfer entropy~\cite{Yada_2022}.
For large $\Delta t$, this scenario corresponds to performing the single-measurement scenario discussed above twice. For small $\Delta t$, the classical protocol does not benefit from the second measurement. In contrast, the extracted work in the quantum protocol can be increased already for small $\Delta t$ since the energy source is the measurement rather than the environment.
%The letter quantity, which bounds the average entropy production by the virtue of Eq.~\eqref{eq: SL Sagawa}, is the quantum-classical transfer entropy~\cite{Yada_2022}, a generalization of the quantum-classical mutual information to the multiple and continuous measurement in quantum systems, see SM for details. %As in the single measurement scenario, the extracted work in the quantum protocol is larger than in the classical case.  

%\begin{equation}
%\label{eq: QC te}
%    \langle I_{QC} \rangle = \sum_{i=0}^{M-1} \sum_{Y_i} P[Y_i] \mathcal{I}(\op{\rho}_{t_{m_{i+1}}}^{Y_{i}}),
%\end{equation}
%where
%\begin{equation}
%    \mathcal{I}(\op{\rho}_{t_{m_{i+1}}}^{Y_{i}}) = \left(S\left(\op{\rho}_{t_{m_{i+1}}}^{Y_{i}} \right) - \sum_{y_{i+1}} P[y_{i+1}|Y_i] S\left( \op{\rho}_{t_{m_{i+1}}}^{Y_{i+1}} \right) \right).
%\end{equation}

\emph{Continuous feedback control--}
Finally, we illustrate our FT and the second law in a continuous feedback protocol. We consider a qubit with a Hamiltonian $\op{H}(t) = \frac{\omega}{2} \op{\sigma}_z +  \chi \op{\sigma}_x \cos{(\Omega t)}$. The effect of the environment is characterized by the same jump operators as above, $\hat{L}_\pm$. The continuous quantum measurement is realized with the measurement operator $\op{M}_1$ from the classical scenario at a rate $\kappa_{\rm m}$. In each infinitesimal time step of duration $\delta t$, the measurement operator is applied with probability $\kappa_{\rm m} \delta t$. The time evolution will thus be interrupted by measurements of the excited state occurring with rate $\kappa_{\rm m}$ and error probability $\epsilon$. Upon the detection of the excited state, a unitary transformation given by $\op{\sigma}_x$ is applied to extract energy from the system. We note that an analogous example was investigated in Ref.~\cite{Yada_2022}. The ensemble averages $\langle e^{-\sigma} \rangle $ and $\langle e^{-(\sigma - \sigma_{\rm cg})} \rangle$ are presented in Fig.~\ref{fig:ContinuousToyModel}~(a), and the second law of thermodynamics is illustrated in Fig.~\ref{fig:ContinuousToyModel}~(b), where the plots are obtained using numerical simulations (as above, the reference state is the thermal state of the final Hamiltonian). Our results are consistent with the FT given in Eq.~\eqref{eq: FT main}, whereas the ordinary FT breaks down. Furthermore, as the strength of the measurement $\kappa_{\rm m}$ grows, the coarse-grained entropy $\langle \sigma_{\rm cg} \rangle$ provides an increasingly tighter bound in comparison with the quantum-classical transfer entropy, $\langle I_{\rm te} \rangle$. This demonstrates that our second law of information thermodynamics~\eqref{eq: SL main} is suitable to bound the entropy production in quantum systems, even in the limit of continuous and strong measurements.

\emph{Conclusions--}
We have derived a FT and a second law of information thermodynamics for quantum system under feedback control. Our results hold for arbitrary open quantum systems and do not rely on any weak coupling or Markovian approximations. The information term appearing in the FT is provided by the coarse-grained entropy, which can be understood as the entropy production inferable from the measurement outcomes. In contrast to the quantum-classical transfer entropy, coarse-grained entropy can thus be computed from the measurement outcomes alone. We have illustrated our findings on a feedback controlled qubit and we found that our second law applies even to the limit of continuous and strong measurements, where the quantum-classical transfer entropy does no longer provide a useful bound on the entropy production. Our work thus fills an important gap for deriving quantum FTs and second laws for arbitrary measurement and feedback schemes.  

Remarkably, the coarse-grained entropy appears in a number of FTs~\cite{Kawai_2007,rahav_2007,gomez_2008,kawaguchi_2013}, notably under imperfect detection (without feedback)~\cite{ferri_2024} and for classical systems in the presence of measurement and feedback~\cite{Sagawa_2012,Potts_2018}. The fact that Eq.~\eqref{eq: detailed FT main} holds across the classical and the quantum regime, and that similar FTs hold in other scenarios, implies that the coarse-grained entropy is a central quantity in stochastic thermodynamics. We believe this is due to its operational character: the coarse-grained entropy provides the part of the entropy production that can be inferred from the available information.

%While the same quantity provides an information term in the \textit{classical} regime~\cite{Potts_2018}, remarkably, here we show that it is also the case in the \textit{quantum} regime, where measurement backaction disrupts dynamics of the system.}

\emph{Acknowledgements--}
P.P.P. and K.P. acknowledge funding from the Swiss National Science Foundation (Eccellenza Professorial Fellowship PCEFP2\_194268).

\bibliography{refs}

%\end{document}
%%%%%%%%%% Merge with supplemental materials %%%%%%%%%%
%\pagebreak

\widetext
\pagebreak
\begin{center}
\textbf{\large Supplemental Material: Quantum Fluctuation Theorem for Arbitrary Measurement and Feedback Schemes}
\end{center}
\begin{center}
Kacper Prech$^{1,*}$ and Patrick P. Potts$^1$
\end{center}
\begin{center}
$^1$\textit{Department of Physics and Swiss Nano Science Institute, University of Basel, Klingelbergstrasse 82, 4056 Basel, Switzerland}
\flushleft
\hspace{1.9cm}$^*$ kacper.prech@unibas.ch
\end{center}

%%%%%%%%%% Merge with supplemental materials %%%%%%%%%%
%%%%%%%%%% Prefix a "S" to all equations, figures, tables and reset the counter %%%%%%%%%%
\setcounter{equation}{0}
\setcounter{figure}{0}
\setcounter{table}{0}
\setcounter{page}{1}
\makeatletter
\renewcommand{\theequation}{S\arabic{equation}}
\renewcommand{\thefigure}{S\arabic{figure}}
\renewcommand{\thetable}{S\arabic{table}}
%\renewcommand{\bibnumfmt}[1]{[S#1]}
%\renewcommand{\citenumfont}[1]{S#1}
%%%%%%%%%% Prefix a "S" to all equations, figures, tables and reset the counter %%%%%%%%%%

In this supplementary material we provide derivations of the detailed fluctuation theorem (FT)~\eqref{eq: detailed FT main}, the integral FT~\eqref{eq: FT main}, and the second law of information thermodynamics~\eqref{eq: SL main} presented in the main text. We also provide supplementary calculations on the examples used to illustrate our findings in the main text. The supplementary material is structured as follows. In Sec.~I we provide derivations of our main results for the general scenario, where the joint quantum state of system and environment is modeled by unitary time-evolution interrupted by measurements. In Sec.~II we focus on scenarios where the reduced system is modeled with a Lindblad master equation.
Sec.~III contains details about the quantum-classical transfer entropy, the information term that the coarse-grained entropy is compared to in the main text.
In Sec.~IV we discuss the qubit model with discrete measurement and feedback. The continuous measurement and the corresponding qubit model are discussed in Sec.~V. Finally, in Sec.~VI we provide details about the measurements in the backward experiment.

\section{I. Unitary evolution}
\subsection{Forward experiment}
The total Hamiltonian of the system and the reservoir has already been introduced in the main text of the letter, but we restate it here for convenience. It is given by
\begin{equation}
    \op{H}_{\rm tot}(\lambda_t^{Y_n}) = \op{H}(\lambda_t^{Y_n}) +   \op{H}_{\rm R} + \op{V}(\lambda_t^{Y_n}),
\end{equation}
where $\op{H}(\lambda_t^{Y_n})$ is the Hamiltonian of the system, $\op{H}_{\rm R} = \sum_E E \proj{E}$ is the Hamiltonian of the reservoir, and $\op{V}(\lambda_t^{Y_n})$ is the coupling Hamiltonian. At the beginning of the experiment, the joint quantum state is assumed to be of the form
\begin{equation}
    \op{\varrho}_0 = \op{\rho}_0 \otimes \op{\tau}_{\rm R},
\end{equation}
where $\op{\rho}_0 = \sum_a p_a \proj{a}$ is the initial state of the system, and $\op{\tau}_{\rm R} = e^{- \beta \op{H}_{\rm R}} /Z_{\rm R} $ is the thermal state of the reservoir. The probability of the trajectory $\Gamma = \{Y, \gamma \}$, where $\gamma = \{ a , E_0, f_Y, E_\tau \}$, is given by
\begin{equation}
    P[\Gamma] = p_a p_{E_0} \text{Tr} \left\{ \proj{f_Y} \otimes \proj{E_\tau}  \left( \prod_{n = 1}^M \mc{U}_n(Y_n) \mc{M}_n(y_n) \right) \mc{U}_0\proj{a} \otimes \proj{E_0}   \right\},
\end{equation}
where $p_{E} = e^{-\beta E}/Z_{\rm R}$ and $\ket{f_Y}$ is an eigenstate of the reference state $\op{\rho}_{\rm r}^Y = \sum_{f_Y} p_{f_Y} \proj{f_Y}$. The reference state may in general be different for each $Y$(generalizing the scenario considered in the main text). We have also introduced a measurement superoperator
\begin{equation}
    \mc{M}_n(y_n) \op{\varrho} = \op{M}_n(y_n) \otimes \op{I}_{\rm R} \op{\varrho} \op{M}_n^{\dagger}(y_n) \otimes \op{I}_{\rm R}
\end{equation}
and a unitary superoperator
\begin{equation}
    \mc{U}_n(Y_n) \op{\varrho} = \hat{U}_n(Y_n) \op{\varrho} \hat{U}_n^{\dagger}(Y_n),
\end{equation}
where the unitary evolution in between the measurements is described by an operator
\begin{equation}
    \hat{U}_n(Y_n) = \mc{T} e^{-i \int_{t_n}^{t_{n+1}} dt \op{H}_{\rm tot}\left(\lambda_t^{Y_n} \right)},
\end{equation}
where $t_0 = 0$, $t_{M+1} = \tau$, and $\mathcal{T}$ is the time ordering operator.
\subsection{Entropy production}
The entropy production corresponding to the trajectory $\Gamma$ may be written as~\cite{Manzano_2018, Manzano_2015, Funo_2015, Landi_2021, Campisi_2011}
\begin{equation}
\label{eq:suppent}
    \sigma[\Gamma] = -\ln{p_{f_Y}} + \ln{p_a}  -\ln{p_{E_\tau}} + \ln{p_{E_0}} = 
     \Delta S[\Gamma] + \beta Q[\Gamma],
\end{equation}
where $\Delta S[\Gamma] = -\ln{p_{f_Y}} + \ln{p_a} $ and $Q[\Gamma] = E_\tau - E_0$ is the stochastic heat transferred to the reservoir. The expectation value of the stochastic heat is equal to the average energy change of the reservoir
\begin{equation}
\begin{split}
    \langle Q \rangle &=  \sum_\Gamma P[\Gamma] Q[\Gamma]  \\
    &= \sum_{a, E_0, f_Y, E_\tau, Y} (E_\tau - E_0) p_a p_{E_0} \text{Tr}  \left\{ \proj{f_Y} \otimes \proj{E_\tau}  \left( \prod_{n = 1}^M \mc{U}_n(Y_n) \mc{M}_n(y_n) \right)\mc{U}_0 \proj{a} \otimes \proj{E_0}   \right\} \\
    & = \sum_{Y, E_\tau} P[Y] E_\tau \text{Tr} \{\op{I}_S \otimes \proj{E_\tau} \op{\varrho}_\tau^Y \}  - \sum_{E_0} E_0 p_{E_0} \\
    & = \sum_{Y} P[Y]  \text{Tr} \{\op{I}_S \otimes \op{H}_{\rm R} \op{\varrho}_\tau^Y \}  - \text{Tr} \{ \op{H}_{\rm R} \op{\tau}_{\rm R} \}.
\end{split} 
\end{equation}
In the calculation we have used the measurement update rule
\begin{equation}
    \op{\varrho}_{t_{n}}^{Y_{n}} = \frac{\mathcal{M}_{n}(y_{n}) \op{\varrho}_{t_{n}}^{Y_{n-1}} }{P[y_{n}|Y_{n-1}]},
\end{equation}
where $P[y_{n+1}|Y_n] = \text{Tr} \{  \mathcal{M}_{n+1}(y_{n+1}) \op{\varrho}_{t_{n+1}}^{Y_{n}}  \}$.
After performing a similar calculation, we find that the ensemble average of $\Delta S[\Gamma]$ is given by
\begin{equation}
\label{eq: average S unitary}
%\begin{split}
    \langle  \Delta S \rangle  =  \sum_\Gamma P[\Gamma] ( -\ln{p_{f_Y}} + \ln{p_a})  %\\
 %   &= \sum_{a, E_0, f_Y, E_\tau, Y} ( -\ln{p_{f_Y}} + \ln{p_a} ) p_a p_{E_0} \text{Tr} \left\{ \proj{f_Y} \otimes \proj{E_\tau}  \left( \prod_{n = 1}^M \mc{U}_n(Y_n) \mc{M}_n(y_n) \right)\mc{U}_0 \proj{a} \otimes \proj{E_0}   \right\} \\
  %  & = - \sum_{f_Y, Y} P[Y] \ln{p_{f_Y}} \text{Tr} \{\op{I}_S \otimes \proj{f_Y} \op{\varrho}_\tau^Y \}  + \sum_{a} p_a \ln{p_a} \\
     = - \sum_{f_Y, Y} P[Y] \bra{f_Y} \op{\rho}_\tau^Y \ket{f_Y} \ln{p_{f_Y}}   + \sum_{a} p_a \ln{p_a}.
%\end{split} 
\end{equation}

The first choice for the reference state is the average state of the system at the end of the experiment
\begin{equation}
    \hat{\rho}_{\rm r} = \op{\rho }_\tau \equiv \sum_Y P[Y] \op{\rho }_\tau^Y.
\end{equation}
In this case, $\Delta S[\Gamma]$ corresponds to the change of the stochastic von Neumann entropy between the final and the initial states. The average entropy production becomes
\begin{equation}
    \langle \sigma \rangle =   S_{\rm vn}\left( \hat{\rho}_\tau \right) - S_{\rm vn} \left( \hat{\rho}_0 \right)  + \beta \langle Q \rangle,
\end{equation}
where $S_{\rm vn}(\op{\rho}) = - \text{Tr} (\op{\rho} \ln{\op{\rho}} )$ is the von Neumann entropy~\cite{nielsen2010quantum}. Alternatively, we may chose the reference state
\begin{equation}
    \op{\rho}_{\rm r}^Y = e^{-\beta \op{H}(\lambda_\tau^Y)}/Z(\lambda_\tau^Y),
\end{equation}
where $Z(\lambda_\tau^Y) = \text{Tr} \{ e^{-\beta \op{H}(\lambda_\tau^Y)} \}$. We also assume that the initial state of the system is given by a thermal state $\op{\rho}_0 = e^{-\beta \op{H}(\lambda_0)}/Z(\lambda_0)$, where $Z(\lambda_0) = \text{Tr} \{ e^{-\beta \op{H}(\lambda_0)} \}$. In this scenario, the expectation value of the entropy production is given by
\begin{equation}
\label{eq: sigma thermal}
    \langle \sigma \rangle = \sum_Y P[Y] \beta \left(  \text{Tr} \{  \op{H}(\lambda_\tau^Y) \op{\rho}_\tau^Y \} - F(\lambda_\tau^Y) \right)  -  \beta\left( \text{Tr}\{ \op{H}(\lambda_0) \op{\rho}_0 \} - F(\lambda_0)  \right) + \beta \langle Q \rangle,
\end{equation}
where $F(\lambda) = -\frac{1}{\beta} \ln{Z(\lambda)}$. If the Kraus operators commute with the Hamiltonian of the system, measurements do not affect the energy of the system. In this case, when the energy stored in the interaction Hamiltonian is small~\cite{Funo_2015}, the entropy production in Eq.~\eqref{eq: sigma thermal} can be connected to the work $\langle W \rangle$ extracted in the feedback driving, $\langle \sigma \rangle = - \beta \langle W \rangle - \sum_Y P[Y] \beta \left( F(\lambda_\tau^Y) - F(\lambda_0) \right)$. In the general case, measurements may change the energy of the system, and $\langle \sigma \rangle$ will include an additional contribution from the energy change due to the measurement. This energy injection has been referred to as quantum heat~\cite{Elouard_2017_qh}, and provides the basis for measurement driven heat engines~\cite{Elouard_2017_engine}.  
%is the equilibrium free energy. The interpretation of $\sum_Y P[Y] \text{Tr} \{  \op{H}(\lambda_\tau^Y) \op{\rho}_\tau^Y \} - \text{Tr}\{ \op{H}(\lambda_0) \op{\rho}_0 \} + \langle Q \rangle $ as work $\langle W \rangle$ is justified when the coupling Hamiltonian either vanishes at the beginning and the end of the experiment, or when it is is small in comparison with the remaining terms in the total Hamiltonian $\op{H}_{\rm tot}$~\cite{Funo_2015}.

\subsection{Backward experiment}
In the backward experiment, a fixed protocol $\{\xi_t \}$ is applied to the the initial state $\op{\Theta} \op{\rho}_{\rm r}^Y \otimes \op{\tau}_{\rm R} \op{\Theta}^{-1}$. Unlike in the forward experiment, this protocol is \textit{not} determined by measurement outcomes but it is fixed from the start of the backward experiment. In addition, measurements are performed. The probability of the inverse trajectory $\bar{\Gamma} = \{\bar{\gamma}, \bar{Y} \}$, where $\bar{\gamma} = \{f_Y, E_\tau, a, E_0 \}$, given the protocol $\{\xi_t \}$ is then given by
\begin{equation}
\label{eq: P tr}
    P_{\rm tr}[\bar{\Gamma}|\{\xi_t \}] = p_{f_Y} p_{E_\tau} \text{Tr} \left\{ \Theta \proj{a}  \otimes  \proj{E_0} \Theta^{-1}   \bar{\mc{U}}_0(\{\xi_t \}) \left(\prod_{n = M}^1 \bar{\mc{M}}_n(y_n) \bar{\mc{U}}_n(\{\xi_t \})\right) \Theta \proj{f_Y}  \otimes  \proj{E_\tau} \Theta^{-1}   \right\}.
\end{equation}
Here we have introduced
\begin{equation}
    \bar{\mc{U}}_n(\{\xi_t \}) \op{\varrho} = \bar{U}_n(\{\xi_t \}) \op{\varrho} \bar{U}_n^{\dagger}(\{\xi_t \}),
\end{equation}
\begin{equation}
    \bar{U}_n(\{\xi_t \}) = \mc{T} e^{-i \int_{\tau - t_{n+1}}^{\tau - t_{n}} dt \op{\Theta} \op{H}_{\rm tot}\left(\xi_t \right) \op{\Theta}^{-1}},
\end{equation}
and
\begin{equation}
    \bar{\mc{M}}_n(y_n) \op{\varrho} = \op{\Theta} \op{M}^{\dagger}_n(y_n)  \otimes \op{I}_{\rm R} \op{\Theta}^{-1} \op{\varrho} \op{\Theta} \op{M}_n(y_n)  \otimes \op{I}_{\rm R} \op{\Theta}^{-1}.
\end{equation}
In contrast to the forward experiment, we kept the explicit $\{\xi_t \}$ dependence in the unitary steps, since the protocol is not determined by the measurement outcomes.
In order to obtain Eq.~\eqref{eq: FT sigma}, we use the principle of microreversibility for driven quantum systems~\cite{Campisi_2011}
\begin{equation}
\label{eq: Microreversibility}
    \op{\Theta}^{-1} \bar{U}_n (\{\lambda_{\tau - t}^Y \}) \op{\Theta}  = \hat{U}_n^{\dagger}(Y_n).
\end{equation}
This relation can be derived by diving the interval $t_{n+1} - t_n$ into $N$ steps of duration $\delta t$ and using a trotterization
\begin{equation}
    \bar{U}_n (\{\lambda_{\tau - t}^Y \}) = e^{-i \delta t \op{\Theta} \op{H}_{\rm tot}\left(\lambda^Y_{\tau - (\tau - t_n)} \right) \op{\Theta}^{-1}}  \cdot\cdot\cdot  e^{-i \delta t \op{\Theta} \op{H}_{\rm tot}\left(\lambda^Y_{\tau - (\tau - t_{n+1})} \right) \op{\Theta}^{-1}} = e^{-i \delta t \op{\Theta} \op{H}_{\rm tot}\left(\lambda^Y_{t_n} \right) \op{\Theta}^{-1}} \cdot\cdot\cdot  e^{-i \delta t \op{\Theta} \op{H}_{\rm tot}\left(\lambda^Y_{t_{n+1}} \right) \op{\Theta}^{-1}}.
\end{equation}
The time reversal operator is antiunitary, $\op{\Theta} i = -i \op{\Theta}$, which leads to $e^{-i \delta t \op{\Theta} \op{H}_{\rm tot}\left(\lambda^Y_t \right) \op{\Theta}^{-1}} = \op{\Theta} e^{i \delta t \op{H}_{\rm tot}\left(\lambda^Y_t \right) } \op{\Theta}^{-1}$. Upon trotterizing $\op{U}_n$ and taking its Hermitian conjugate, we obtain Eq.~\eqref{eq: Microreversibility}. The principle of microreversibility, together with the cyclic property of the trace, results in
\begin{equation}
    P_{\rm tr}[\bar{\Gamma}|\{\lambda_{\tau - t}^Y \}] = p_{f_Y} p_{E_\tau} \text{Tr} \left\{ \proj{f_Y} \otimes \proj{E_\tau}  \left( \prod_{n = 1}^M \mc{U}_n(Y_n) \mc{M}_n(y_n)\right) \mc{U}_0 \proj{a} \otimes \proj{E_0}   \right\},
\end{equation}
which, together with Eq.~\eqref{eq:suppent}, yields Eq.~\eqref{eq: FT sigma}.

\subsection{Fluctuation theorems and the second law}
The coarse-grained entropy is defined as~\cite{Kawai_2007, Sagawa_2012, Potts_2018}
\begin{equation}
\label{eq: sigma cg def}
    e^{-\sigma_{\rm cg}[Y]} \equiv \sum_\gamma e^{-\sigma[\Gamma]} P[\gamma | Y],
\end{equation}
where $P[\gamma | Y]  = P[\Gamma]/P[Y]$ is the conditional probability for the part of the trajectory that is not accessible by the measurements. The probability to obtain the set of the measurement outcomes $Y$ is given by
\begin{equation}
\label{eq: PY}
\begin{split}
    P[Y] &= \sum_\gamma P[\Gamma] \\
    & = \sum_{a, E_0, f_Y, E_\tau}  p_a p_{E_0} \text{Tr} \left\{ \proj{f_Y} \otimes \proj{E_\tau}  \left( \prod_{n = 1}^M \mc{U}_n(Y_n) \mc{M}_n(y_n) \right) \mc{U}_0 \proj{a} \otimes \proj{E_0}  \right\} \\
    & = \text{Tr} \left\{ \left(\prod_{n = 1}^M \mc{U}_n(Y_n) \mc{M}_n(y_n) \right)\mc{U}_0 \op{\varrho}_0 \right\}.
\end{split}
\end{equation}
The coarse-grained entropy may then be expressed as 
\begin{equation}
\label{eq: cg to m}
    e^{-\sigma_{\rm cg}[Y]} = \frac{1}{P[Y]} \sum_\gamma e^{-\sigma[\Gamma]} P[\Gamma] = \frac{1}{P[Y]} \sum_{\bar{\gamma}} P_{\rm tr}[\bar{\Gamma}|\{\lambda_{\tau - t}^Y \}] = \frac{P_{\rm tr}[\bar{Y}| \{\lambda_{\tau - t}^Y\}]}{P[Y]} ,
\end{equation}
recovering Eq.~\eqref{eq: sigma cg m} in the main text, where to obtain the second equality we used Eq.~\eqref{eq: FT sigma}, and in the third equality we used the definition of $ P_{\rm tr}[\bar{Y}| \{\lambda_{\tau - t}^Y\}] $. The last quantity can be found from
\begin{equation}
        P_{\rm tr}[\bar{Y}| \{\lambda_{\tau - t}^Y\}] =  \text{Tr} \left\{  \left(\prod_{n = M}^1 \bar{\mc{M}}_n(y_n) \bar{\mc{U}}_n \right)\Theta \op{\rho}_{\rm r}^Y  \otimes   \op{\tau}_{\rm R} \Theta^{-1}   \right\} .
\end{equation}
The detailed FT~\eqref{eq: detailed FT main} can then be obtained by combining the full backward probability~\eqref{eq: PB} with Eqs.~\eqref{eq: FT sigma} and~\eqref{eq: sigma cg m}.
The integral FT~\eqref{eq: FT main} follows from
\begin{equation}
    \begin{split}
        \langle e^{-(\sigma - \sigma_{\rm cg})} \rangle &= \sum_\Gamma P[\Gamma] e^{-(\sigma[\Gamma] - \sigma_{\rm cg}[Y])} = \sum_\Gamma P_{\rm B}[\bar{\Gamma}] = \sum_Y \frac{P[Y]}{P_{\rm tr}[\bar{Y}|\{\lambda_{\tau - t}^Y\}]} \sum_{\gamma} P_{\rm tr}[\bar{\Gamma}|\{\lambda_{\tau - t}^Y \}] \\
        & = \sum_Y \frac{P[Y]}{P_{\rm tr}[\bar{Y}|\{\lambda_{\tau - t}^Y\}]} P_{\rm tr}[\bar{Y}|\{\lambda_{\tau - t}^Y\}] = \sum_Y P[Y]\\
        & = \sum_Y \text{Tr} \left\{ \left(\prod_{n = 1}^M \mc{U}_n(Y_n) \mc{M}_n(y_n)\right) \mc{U}_0 \op{\varrho}_0 \right\} = \text{Tr} \{ \op{\varrho}_0 \} = 1, \\
    \end{split}
\end{equation}
where in the second equality we used the detailed FT~\eqref{eq: detailed FT main}, in the third equality we used Eq.~\eqref{eq: PB},
in the fourth inequality we applied the definition of $ P_{\rm tr}[\bar{Y}|\{\lambda_{\tau - t}^Y\}]$,
and in the sixth in equality we used Eq.~\eqref{eq: PY}.

The second law of information thermodynamics~\eqref{eq: SL main} follows from Jensen's inequality, $\langle f(X)\rangle \leq f(\langle X \rangle)$ for a convex function $f(X)$, applied to the integral FT~\eqref{eq: FT main}. The ensemble average of the coarse-grained entropy can be computed using
\begin{equation}
    \langle \sigma_{\rm cg} \rangle = \sum_Y P[Y] \ln{ \frac{P[Y] }{ P_{\rm tr}[\bar{Y}|\{\lambda_{\tau - t}^Y\}] } }.
\end{equation}

\section{II. Lindblad master equation}
\subsection{Forward experiment}
In many situations, the time-evolution of the density matrix of the system $\op{\rho}_t^{Y_n}$ in between the measurements $n$ and $n+1$ may be described using the Lindblad master equation~\eqref{eq:Lindblad forward}, which is restated here for convenience
\begin{equation}
    \frac{d}{dt} \op{\rho}_t^{Y_n} =   -i[\hat{H}(\lambda_t^{Y_n}) , \op{\rho}_t^{Y_n} ] + \sum_j \mathcal{D}[\op{L}_j(\lambda_t^{Y_n})]\op{\rho}_t^{Y_n} = \mathcal{L}(\lambda_t^{Y_n}) \op{\rho}_t^{Y_n}.
\end{equation}
In the small time interval $dt$, the change in the quantum state may be expressed as%~\cite{Manzano_2022, Manzano_2018, Yada_2022, Horowitz_2012, Horowitz_2013, Gong_2016, wiseman2010quantum}
\begin{equation}
\label{eq: dt}
    \op{\rho}_{t + dt}^{Y_n} = \op{\rho}_{t}^{Y_n} -i \left( \op{H}_{\rm eff}(\lambda_t^{Y_n}) \op{\rho}_{t}^{Y_n} - \op{\rho}_{t}^{Y_n}  \op{H}_{\rm eff}^\dagger(\lambda_t^{Y_n}) \right) dt + \sum_j  \op{L}_j(\lambda_t^{Y_n}) \op{\rho}_{t}^{Y_n} \op{L}_j^{\dagger}(\lambda_t^{Y_n}) dt,
\end{equation}
where we have introduced an effective non-Hermitian Hamiltonian
\begin{equation}
\label{eq: NH}
    \op{H}_{\rm eff}(\lambda_t^{Y_n}) = \op{H}(\lambda_t^{Y_n}) - \frac{i}{2} \sum_j \op{L}_j^{\dagger}(\lambda_t^{Y_n}) \op{L}_j(\lambda_t^{Y_n}).
\end{equation}
Equation \eqref{eq: dt} may be unravelled into trajectories. Along a single trajectory, the conditional quantum state $\op{\rho}_{\rm c}$ follows a stochastic master equation
\begin{equation}
\label{eq: Stoch ME }
    d\op{\rho}_{\rm c} = -i \left( \op{H}_{\rm eff}(\lambda_t^{Y_n}) \op{\rho}_{\rm c} - \op{\rho}_{\rm c} \op{H}_{\rm eff}^\dagger(\lambda_t^{Y_n}) \right) dt + \sum_j \text{Tr} \{ \op{L}_j^{\dagger}(\lambda_t^{Y_n}) \op{L}_j(\lambda_t^{Y_n}) \op{\rho}_{\rm c} \}  \op{\rho}_{\rm c} dt  + \sum_j \left( \frac{ \op{L}_j(\lambda_t^{Y_n}) \op{\rho}_{\rm c} \op{L}_j^{\dagger}(\lambda_t^{Y_n}) }{\text{Tr} \{ \op{L}_j^{\dagger}(\lambda_t^{Y_n}) \op{L}_j(\lambda_t^{Y_n}) \op{\rho}_{\rm c} \} }- \op{\rho}_{\rm c} \right) dN_j,
\end{equation}
where $dN_j$ are random variables taking a value $1$ if the jump $j$ occurs and $0$ otherwise, obeying the relation $dN_j dN_{j'} = \delta_{j, j'} dN_j$. The probability for a jump to occur is given by $p_c^j= \text{Tr} \{ \op{L}_j^{\dagger}(\lambda_t^{Y_n}) \op{L}_j(\lambda_t^{Y_n}) \op{\rho}_{\rm c} \}  dt$. %Measurement selective quantum state $\op{\rho}_t^{Y_n} = \langle \op{\rho}_{\rm c}  \rangle$ is the ensemble average of $ \op{\rho}_{\rm c} $ over the quantum jump trajectories of the system $\gamma$ up to the time $t$. Measurement selective expectation values of the random variables $dN_j$ averaged over $\gamma$ up to the time $t$ are given by $\langle dN_j \rangle = \langle p^j_{\rm c} \rangle = \text{Tr} \{ \op{L}_j(\lambda_t^{Y_n})^{\dagger} \op{L}_j(\lambda_t^{Y_n}) \langle \op{\rho}_{\rm c} \rangle \}  dt = \text{Tr} \{ \op{L}_j(\lambda_t^{Y_n})^{\dagger} \op{L}_j(\lambda_t^{Y_n}) \op{\rho}_t^{Y_n} \}  dt$.

The probability density of the trajectory $\Gamma$ corresponding to the unravelling~\eqref{eq: Stoch ME } reads
\begin{equation}
\label{eq: P Gamma Lindblad}
    P[\Gamma] = p_a \text{Tr} \left\{ \proj{f_Y}  \mc{T} \left\{  \mc{U}_{\rm eff}(\tau,0) \prod_{n = 1}^M \mc{M}_n(y_n) \prod_{k = 1}^K \mc{J}_k(j_k, s_k) \right\}   \proj{a}  \right\}  , 
%\end{equation}
\end{equation}
where
\begin{equation}
\label{eq: Measurment}
    \mc{M}_n(y_n) \op{\rho} = \op{M}_n(y_n)  \op{\rho} \op{M}^{\dagger}_n(y_n)
\end{equation}
is the superoperator corresponding to the measurement at the time $t_n$,
\begin{equation}
\label{eq: Jump}
    \mc{J}_k(j_k, s_k) \op{\rho} = \hat{L}_{j_{k}}(\lambda^{Y_n}_{s_{k}})  \op{\rho} \hat{L}_{j_{k}}^\dagger(\lambda^{Y_n}_{s_{k}})
\end{equation}
is a superoperator describing the $k$-th jump $j_k$ at the time $s_k$, and $\mc{U}_{\rm eff}(\tau, 0)$ an effective nonunitary superoperator that governs the time-evolution of the system from $t = 0$ to $t = \tau$ in between measurements and jumps. It is given by
\begin{equation}
\label{eq: U super eff}
    \mc{U}_{\rm eff}(t'', t') \op{\rho} = \hat{U}_{\rm eff}(t'', t') \op{\rho} \hat{U}_{\rm eff}^\dagger(t'', t'),
\end{equation} 
where
\begin{equation}
\label{eq: U eff}
    \hat{U}_{\rm eff}(t'', t') = \mc{T} e^{-i \int_{t'}^{t''} dt \op{H}_{\rm eff}(\lambda_t^{Y_n})}.
\end{equation}
Note that $P[\Gamma]$ is a probability density. To obtain the probability that each jump in the trajectory $\Gamma$ occurs within the time-window $ds$ around the times $s_k$, it needs to be multiplied by $(ds_k)^K$. %Between consecutive measurements or jumps at $t'$ and $t''$, corresponding time-evolution superoperator $\mc{U}_{\rm eff}(t'', t') \op{\rho} = U_{\rm eff}(t'', t') \op{\rho} U_{\rm eff}(t'', t')^\dagger$ is generated with a nonunitary $U_{\rm eff}(t'', t') = \mc{T} e^{-i \int_{t'}^{t''} dt \op{H}_{\rm eff}(\lambda_t^{Y_n})}$. 
The time ordering superoperator $\mc{T} \{ \cdot \}$ in Eq.~\eqref{eq: P Gamma Lindblad} ensures that all the superoperators act in the correct order. We use this notation instead of explicitly writing all the effective nonunitary superoperators in between jumps and measurements to simplify the notation. For example, in the case of a single measurement and for a trajectory with a single jump occurring after the measurement,
\begin{equation}
    \mc{T} \left\{  \mc{U}_{\rm eff}(\tau,0)  \mc{M}_1(y_1) \mc{J}_1(j_1, s_1) \right\}   \proj{a} = \mc{U}_{\rm eff}(\tau, s_1) \mc{J}_1(j_1, s_1) \mc{U}_{\rm eff}(s_1, t_1) \mc{M}_1(y_1)  \mc{U}_{\rm eff}(t_1, 0) \proj{a}.
\end{equation}

%\begin{equation}
%    P[\Gamma] = p_a  \bra{f_Y}  \prod_{n = 1}^M  \mathcal{J}_n \mathcal{M}_n(y_n)  \mathcal{J}_0 \proj{a} \prod_{k = 1}^K ds_k   \ket{f_Y}, 
%\end{equation}
%where $\mathcal{J}_n \op{\rho} = \mathbb{J}_n \op{\rho} \mathbb{J}_n^{\dagger} $,
%\begin{equation}
%    \mc{M}_n(y_n) \op{\rho} = \op{M}_n(y_n)  \op{\rho} \op{M}_n(y_n)^{\dagger},
%\end{equation}
%and
%\begin{equation}
%    \mathbb{J}_n = U_{\rm eff}(t_{n+1} , s_{K_{n+1}}) L_{j_{K_{n+1}}}(\lambda^Y_{s_{K_{n+1}}}) \prod_{k = K_{n}+1}^{K_{n+1} - 1} U_{\rm eff}(s_{k+1} , s_k) L_{j_k}(\lambda^Y_{s_k})U_{\rm eff}(s_{K_n +1 } , t_n).
%\end{equation}
%Here $K_n$ denotes the number of Lindblad jumps that take place up to the measurement $n+1$, $K_0 = 0$, and $K_{M+1} = K$. The supeoperators $\mathcal{J}_n$ describe the evolution in between the measurement $n$ and $n+1$. We have also introduced an effective non-unitary $U_{\rm eff}(t'', t')$ that is generated with the effective non-hermitian Hamiltonian $\op{H}_{\rm eff}(\lambda_t^{Y_n})$ and describes the evolution in between two consecutive jumps occurring at times $t'$ and $t''$.

The entropy production along the trajectory reads~\cite{Manzano_2018}%, Horowitz_2012, Horowitz_2013, Gong_2016, Yada_2022}
\begin{equation}
\label{eq: sigma Lindblad}
    \sigma[\Gamma] = \Delta S [\Gamma] + \beta Q[\Gamma],
\end{equation}
where $\Delta S [\Gamma] = -\ln{p_{f_Y}} + \ln{p_a}$ and $Q[\Gamma] = \sum_{k = 1}^K q_{j_k}(\lambda^{Y_n}_{s_k}) = \sum_j \int_0^\tau dN_j q_j(\lambda_t^{Y_n}) $ is the stochastic heat.
For the expectation value of the change in the system entropy we recover Eq.~\eqref{eq: average S unitary} as expected. The ensemble average of the heat transferred to the bath is given by
\begin{equation}
%\begin{split}
    \langle Q \rangle  = \sum_{\Gamma} P[\Gamma] \sum_j \int_0^\tau dN_j q_j(\lambda_t^{Y_n}) %= \sum_j \sum_Y P[Y] \int_0^\tau \langle dN_j \rangle q_j(\lambda_t^{Y_n}) \\& 
    =  \sum_Y P[Y] \int_0^\tau \sum_j q_j(\lambda_t^{Y_n}) \text{Tr} \{  \op{L}_j(\lambda_t^{Y_n}) \op{\rho}_t^{Y_n} \op{L}_j^{\dagger}(\lambda_t^{Y_n}) \} dt . %= \sum_{n = 0}^M \sum_{Y_n} P[Y_n] \int_{t_{n}}^{t_{n+1}}  q_j(\lambda_t^{Y_n}) \text{Tr} \{  \op{L}_j(\lambda_t^{Y_n}) \op{\rho}_t^{Y_n} \op{L}_j(\lambda_t^{Y_n})^{\dagger} \} dt.
%\end{split}
\end{equation}
This equation is obtained by using that the probability for a jump $j$ at time $t$, averaged over all previous quantum jumps (but not measurement outcomes) is given by $\text{Tr} \{  \op{L}_j(\lambda_t^{Y_n}) \op{\rho}_t^{Y_n} \op{L}_j^{\dagger}(\lambda_t^{Y_n}) \} dt$.

\subsection{Backward experiment}
In the backward experiment, a fixed protocol $\{ \xi_t \}$ is applied starting from an initial state $\op{\Theta} \op{\rho}_{\rm r}^Y \op{\Theta}^{-1} $. The probability density for the trajectory $\bar{\Gamma}$ is given by
\begin{equation}
\label{eq: P tr Lindblad}
    P_{\rm tr}[\bar{\Gamma}|\{\xi_t \}] = p_{f_Y} \text{Tr} \left\{ \op{\Theta} \proj{a} \op{\Theta}^{-1} \mc{T} \left\{ \bar{\mc{U}}_{\rm eff}(\tau, 0) \prod_{n = 1}^M \bar{\mathcal{M}}_{n}(y_n)  \prod_{k = 1}^K \bar{\mc{J}}_k(\bar{s}_k, \bar{j}_k) \right\}   \op{\Theta} \proj{f_Y} \op{\Theta}^{-1}  \right\}, 
\end{equation}
where
\begin{equation}
\label{eq: Measurment B}
    \bar{\mathcal{M}}_{n}(y_{n}) \op{\rho} = \op{\Theta} \op{M}^{\dagger}_{n}(y_{n}) \op{\Theta}^{-1} \op{\rho} \op{\Theta} \op{M}_{n}(y_{n}) \op{\Theta}^{-1}
\end{equation}
is the measurement superoperator of the measurement performed at the time $\tau - t_n$,
\begin{equation}
\label{eq: Jump B}
    \bar{\mc{J}}_k(\bar{s}_k, \bar{j}_k) \op{\rho}= \op{\Theta} \op{L}_{\bar{j}_k} (\xi_{\bar{s}_k}) \op{\Theta}^{-1} \op{\rho} \op{\Theta} \op{L}_{\bar{j}_k}^{\dagger} (\xi_{\bar{s}_k}) \op{\Theta}^{-1},
\end{equation}
is the superoperator of the $k$-th jump occurring at the time $\bar{s}_k = \tau -  s_{k}$ with $\bar{j}_k = \tilde{j}_{k}$, and $\bar{\mc{U}}_{\rm eff}(\tau, 0)$ is an effective nonunitary superoperator describing the time-evolution in between measurements and jumps. Its action is implemented according to Eq.~\eqref{eq: U super eff}, but with a nonunitary operator $\bar{U}_{\rm eff}(t'', t') $ generated with the non-Hermitian Hamiltonian $\op{\Theta} \op{H}_{\rm eff}^\dagger(\xi_t) \op{\Theta}^{-1}=\op{\Theta} \op{H}(\xi_t) \op{\Theta}^{-1} - \frac{i}{2}\sum_j \op{\Theta} \op{L}_j^{\dagger}(\xi_t) \op{L}_j(\xi_t) \op{\Theta}^{-1} $in a complete analogy to Eq.~\eqref{eq: U eff}. This results in the microreversibility condition $\op{\Theta}^{-1} \bar{U}_{\rm eff}(t'', t') \op{\Theta} = \hat{U}_{\rm eff}^{\dagger}(\tau - t', \tau - t'') $~\cite{Gong_2016, Horowitz_2013}. Moreover, we have $L_{\bar{j}_k} = L_{\tilde{j}_{k}} = e^{-\beta q_{k}/2} L_{j_{k}}^{\dagger}$. Using these relations together with the cyclic property of the trace yields
\begin{equation}
     P_{\rm tr}[\bar{\Gamma}|\{\lambda_{\tau - t}^Y \}] = p_{f_Y} \prod_{k = 1}^K e^{-\beta q_{j_k}(\lambda^Y_{s_k}) } \text{Tr} \left\{ \proj{f_Y}  \mc{T} \left\{  \mc{U}_{\rm eff}(\tau,0) \prod_{n = 1}^M \mc{M}_n(y_n) \prod_{k = 1}^K \mc{J}_k(j_k, s_k) \right\}   \proj{a}  \right\},
\end{equation}
which leads to Eq.~\eqref{eq: FT sigma}.

\subsection{Fluctuation Theorems}
The coarse-grained entropy can still be expressed by Eq.~\eqref{eq: sigma cg m}
\begin{equation}
\label{eq: cg Lindblad}
    e^{-\sigma_{\rm cg}[Y]} = \int \mathfrak{D}[\gamma] e^{-\sigma[\Gamma]} P[\gamma|Y] = \frac{1}{P[Y]} \int \mathfrak{D}[\gamma] e^{-\sigma[\Gamma]} P[\Gamma]
    =  \frac{1}{P[Y]} \int \mathfrak{D}[\bar{\gamma}]  P_{\rm tr}[\bar{\Gamma}|\{\lambda_{\tau - t}^Y \}] 
    =  \frac{P_{\rm tr}[\bar{Y}|\{\lambda_{\tau - t}^Y \}]}{P[Y]},
\end{equation}
where in place of $\sum_\gamma$ we have introduced a path integral
\begin{equation}
    \int \mathfrak{D}[\gamma] = \sum_{ f_Y, a} \sum_{K = 0}^{\infty} \prod_{k = 1}^K \sum_{\{j_k\}}  \int_0^{s_{k+1}}  ds_k.
\end{equation}
 The probability of obtaining the set of the measurement outcomes $Y$ can be expressed as
\begin{equation}
    P[Y] = \int \mathfrak{D}[\gamma] P[\Gamma] = \text{Tr} \left\{ \prod_{n = 1}^M \left( \mathcal{M}_n(y_n) \mathcal{T} e^{\int_{t_{n-1}}^{t_n} \mathcal{L}(\lambda_t^{Y_n}) dt}  \right) \op{\rho}_0 \right\},
\end{equation}
and $P_{\rm tr}[\bar{Y}|\{\lambda_{\tau - t}^Y \}]$ can be computed analogously.
%where we used Eq.~\eqref{eq: relation}.
As is the previous section, the detailed FT [see Eq.~\eqref{eq: detailed FT main}] follows from Eqs.~\eqref{eq: PB}, \eqref{eq: FT sigma}, and~\eqref{eq: cg Lindblad}.
The integral FT, given in Eq.~\eqref{eq: FT main}, is obtained as follows:
\begin{equation}
\label{eq: Int FT derivation Lind}
\begin{split}
     \langle e^{-(\sigma - \sigma_{\rm cg})} \rangle =& \sum_Y \int \mathfrak{D}[\gamma] P[\Gamma] e^{-(\sigma[\Gamma] - \sigma_{\rm cg}[Y])}  =  \sum_{Y} \int \mathfrak{D}[\bar{\gamma}] P_{\rm B}[\bar{\Gamma}|\{\lambda_{\tau - t}^Y \}] \\
     = & \sum_{Y} \frac{P[Y]}{P_{\rm tr}[\bar{Y}|\{\lambda_{\tau - t}^Y \}]} \int \mathfrak{D}[\bar{\gamma}] P_{\rm tr}[\bar{\Gamma}|\{\lambda_{\tau - t}^Y \}] =   \sum_{Y} \frac{P[Y]}{P_{\rm tr}[\bar{Y}|\{\lambda_{\tau - t}^Y \}]}  P_{\rm tr}[\bar{Y}|\{\lambda_{\tau - t}^Y \}] = 1,
\end{split}
\end{equation}
where to obtain the second equality and the third equality, we have used Eqs.~\eqref{eq: detailed FT main} and~\eqref{eq: PB}, respectively.

\section{III. Quantum-classical transfer entropy}

The quantum-classical transfer entropy $\langle I_{\rm te} \rangle$, which was recently introduced in Ref.~\cite{Yada_2022}, provides another second law of information thermodynamics for multiple and continuous measurement [cf. Eq.~\eqref{eq:FT intro}]. This quantity is defined as
\begin{equation}
\label{eq: QC te}
    \langle I_{\rm te} \rangle \equiv \sum_{n=0}^{M-1} \sum_{Y_n} P[Y_n] \mathcal{I}(\op{\rho}_{t_{n+1}}^{Y_n}),
\end{equation}
with
\begin{equation}
    \mathcal{I}(\op{\rho}_{t_{n+1}}^{Y_{n}}) = \left(S_{\rm vn}\left(\op{\rho}_{t_{n+1}}^{Y_{n}} \right) - \sum_{y_{n+1}} P[y_{n+1}|Y_n] S_{\rm vn}\left( \op{\rho}_{t_{n+1}}^{Y_{n+1}} \right) \right),
\end{equation}
where
\begin{equation}
\label{eq: update Lindblad}
    \op{\rho}_{t_n}^{Y_{n}} = \frac{ \mathcal{M}_n(y_n)\op{\rho}_{t_n}^{Y_{n-1}} }{P[y_n|Y_{n-1}] },
\end{equation}
is the post measurement updated quantum state, and $P[y_n|Y_{n-1}] = \text{ Tr} \{ \mathcal{M}_n(y_n)\op{\rho}_{t_n}^{Y_{n-1}} \} $. 
If only a single measurement is performed ($M = 1$), the quantum-classical transfer entropy becomes the quantum-classical mutual information,
which appears in Ref.~\cite{Sagawa_2008}.  In classical systems, the quantum-classical transfer entropy reduces to the transfer entropy~\cite{Sagawa_2012},
whereas the quantum-classical mutual information reduces to the mutual information~\cite{Sagawa_2010}. 

From the form of the quantum-classical transfer entropy it can be noticed that its computing requires both knowledge of the measurement outcomes used for feedback and the state of the system before and after each measurement. The latter is not directly accessible in the experiment. This is unlike the coarse-grained entropy, which can be obtained solely from the statistics of measurement outcomes, making it experimentally accessible quantity.

To obtain a stochastic form of the quantum-classical transfer entropy, $I_{\rm te}$, the authors of Ref.~\cite{Yada_2022} consider a so-called alternative quantum jump unravelling, where in each time step $\delta t$ both bath jump and feedback measurement jump may happen. This is explained in more detail in Sec.~V. To define the stochastic $I_{\rm te}$, fictitious measurements onto eigenstates of a pre- and post-measurement density matrix are introduced and incorporated for each quantum trajectory. This is not necessary for a stochastic form of the coarse-grained entropy.

%\begin{equation}
%\begin{split}
%\op{\rho}_{t_{n+1}}^{Y_{n+1}} =& V^{Y_n} \op{\rho}_{t_{n+1}}^{'Y_{n+1}} (V^{Y_n})^{\dagger}, \\
%    \op{\rho}_{t_{n+1}}^{'Y_{n+1}} =& \op{\rho}_{t_n}^{Y_n} -i [\op{H}_{\rm eff}(\lambda_{t_n}^{Y_n}), \op{\rho}_{t_n}^{Y_n}] \delta t + \sum_j \text{Tr} \{ \op{L}_j(\lambda_{t_n}^{Y_n})^{\dagger} \op{L}_j(\lambda_{t_n}^{Y_n}) \op{\rho}_{t_n}^{Y_n} \}  \op{\rho}_{t_n}^{Y_n} \delta t  
%    +  \sum_j \left( \frac{ \op{L}_j(\lambda_{t_n}^{Y_n}) \op{\rho}_{t_n}^{Y_n} \op{L}_j(\lambda_{t_n}^{Y_n})^{\dagger} }{\text{Tr} \{ \op{L}_j(\lambda_{t_n}^{Y_n})^{\dagger} \op{L}_j(\lambda_{t_n}^{Y_n}) \op{\rho}_{t_n}^{Y_n} \}}- \op{\rho}_{t_n}^{Y_n} \right) \delta N_j \\
%    +& \sum_y \text{Tr}\{ \op{M}_n(y) \op{\rho}_{t_n}^{Y_n} \op{M}_n(y)^\dagger \}\op{\rho}_{t_n}^{Y_n} \delta t - \frac{1}{2} \sum_y \{ \op{M}(y)^{\dagger} \op{M}(y), \op{\rho}_{t_n}^{Y_n} \} \op{\rho}_{t_n}^{Y_n} \delta t + \sum_y  \left( \frac{ \op{M}(y) \op{\rho}_{t_n}^{Y_n} \op{M}(y)^{\dagger} }{\text{Tr} \{ \op{M}(y)^{\dagger} \op{M}(y) \op{\rho}_{t_n}^{Y_n} \}}- \op{\rho}_{t_n}^{Y_n} \right) \delta N_y.
%\end{split}
%\end{equation}

\section{IV. Qubit model}

\subsection{Single measurement and feedback}

The system with the Hamiltonian $\op{H} = \frac{\omega}{2} \op{\sigma}_z$ is initially in the quantum state
\begin{equation}
\label{eq:instatesm}
    \op{\rho}_0 = p_1 \proj{1} + p_0 \proj{0},
\end{equation}
where $p_1 = e^{-\beta \omega}/(1+ e^{-\beta \omega})$ and $p_0 = 1/(1+ e^{-\beta \omega})$. In the classical protocol, the state is measured in the computational basis, with corresponding probabilities
\begin{equation}
\label{eq:singpy}
    \begin{split}
        P[0] =& (1- \epsilon) p_0 + \epsilon p_1, \\
        P[1] =& (1- \epsilon) p_1 + \epsilon p_0. \\
    \end{split}
\end{equation}
After the measurement, the density matrix of the system conditioned on the outcome is given by
\begin{equation}
\label{eq: post measurement classical}
    \begin{split}
        \op{\rho}^0 =& \left(p_1 \epsilon \proj{1} + p_0 (1-\epsilon) \proj{0} \right)/P[0],\\
        \op{\rho}^1 =& \left(p_0 \epsilon \proj{0} + p_1 (1- \epsilon) \proj{1}\right)/P[1].\\
    \end{split}
\end{equation}
After the measurement, a unitary operator $\op{U}_Y$ is applied to $\op{\rho}^Y$. For $Y=1$, this swaps the states while for $Y=0$, it does nothing (see main text). Then, the system thermalizes again and reaches $\op{\rho}_\tau = \op{\rho}_0$.

In the backward experiment, $Y$ is chosen randomly from the distribution given in Eq.~\eqref{eq:singpy}. For a given $Y$, the unitary operator $\op{U}_Y^{\dagger}$ is applied to the state $\hat{\rho}_\tau$ and subsequently, the the state is measured in the computational basis with corresponding probabilities
\begin{equation}
        P_{\rm tr}[0|\op{U}_0^{\dagger}] = P_{\rm tr}[1|\op{U}_1^{\dagger}] = (1- \epsilon) p_0 + \epsilon p_1.
\end{equation}
The coarse-grained entropy can be computed as
\begin{equation}
\label{eq: cg calculation}
    \sigma_{\rm cg}[Y] = \ln{\left(P[Y]/P_{\rm tr}[Y|\op{U}_Y^\dagger] \right)} .
\end{equation}

\begin{table}[h]
    \centering
        \begin{tabular}{|c | c | c | c | c| c|c|} 
 \hline
 $a$ & $Y$ & $f$ & $q$ & $ P[\Gamma]$ & $e^{-\sigma}$ & $P_{\rm tr}[\bar{\Gamma} | \op{U}_Y^\dagger ]$  \\  
 \hline\hline
 0 & 0 & 0 & 0 & $p_0 (1-\epsilon)p_0$ & $1$ & $p_0 p_0 (1-\epsilon)$ \\ 
 \hline
 1 & 0 & 0 & $\omega$ & $p_1 \epsilon p_0$  & $\frac{p_0}{p_1} e^{-\beta \omega}$ & $p_0 p_1 \epsilon$\\ 
 \hline
 0 & 0 & 1 & $-\omega$ & $p_0 (1- \epsilon) p_1$ & $\frac{p_1}{p_0} e^{\beta \omega}$ & $p_1 p_0 (1-\epsilon)$\\
 \hline
 1 & 0 & 1 & 0 & $ p_1 \epsilon p_1$ & $ 1 $ & $p_1 p_1 \epsilon$\\
 \hline
 0 & 1 & 0 & $\omega$ & $p_0 \epsilon p_0$ & $e^{-\beta \omega}$ & $p_0 p_1 \epsilon$\\ 
 \hline
 1 & 1 & 0 & 0 & $p_1 (1-\epsilon) p_0$  & $\frac{p_0}{p_1} $ & $p_0 p_0 (1-\epsilon)$\\ 
 \hline
 0 & 1 & 1 & 0 & $p_0 \epsilon p_1$ & $\frac{p_1}{p_0}$ & $p_1 p_1 \epsilon$\\ 
 \hline
 1 & 1 & 1 & $-\omega$ & $p_1 (1-\epsilon) p_1$ & $e^{\beta \omega}$ & $p_1 p_0 (1-\epsilon)$\\
 \hline
\end{tabular}
    \caption{The qubit model with a single classical measurement. Table shows all the combinations of $\Gamma = \{a, Y, f \}$, corresponding stochastic heat $q$, probability $P[\Gamma]$, corresponding stochastic entropy production $\sigma$, and probability $P_{\rm tr}[\bar{\Gamma} |\op{U}_Y^\dagger]$.}
    \label{tab: Clas}
\end{table}

The trajectories discussed in the main text either involve the energies of the environment at the beginning and the end of the protocol (for unitary evolution) or the jumps induced by the environment (for Lindblad master equations). Here, we are not interested in the details of the final thermalization step and we thus group together all trajectories that result in the same heat ($q$) exchanged with the environment. In this case, the trajectory is fully determined by $\Gamma=\{a, Y, f \}$. In Tab.~\ref{tab: Clas} we list all different trajectories, providing the corresponding probabilities as well as the stochastic entropy production. The probability of the trajectory $\Gamma $ can be computed as
\begin{equation}
    P[\Gamma] =  p_a p_f  \text{Tr} \left\{ \op{M}_Y \proj{a} \op{M}_Y^\dagger \right\} =  p_a p_f\left(\delta_{Y,a}(1-\epsilon)+\epsilon(1-\delta_{Y,a})\right).
\end{equation}
From Tab.~\ref{tab: Clas}, we can verify our integral FT $\langle e^{-(\sigma - \sigma_{\rm cg})} \rangle = 1$. In addition, we provide the probability of the inverse trajectory $\Gamma$ when we apply the inverse protocol, which can be found using
\begin{equation}
    P_{\rm tr}[\bar{\Gamma} | \op{U}_Y^\dagger ] = p_f \text{Tr} \left\{ \proj{a} \op{M}_Y^\dagger \op{U}_Y^\dagger \op{\rho}_\tau \op{U}_Y \op{M}_Y  \right\} = p_f \left( \delta_{a, Y}  p_0(1-\epsilon) + (1-\delta_{a, Y}) p_1 \epsilon \right).
\end{equation}
The full backward probability can be computed as $P_{\rm B}[\bar{\Gamma}] = P_{\rm tr}[\bar{\Gamma} | \op{U}_Y^\dagger ] P[Y]/P_{\rm tr}[Y | \op{U}_Y^\dagger ]$ [cf. Eq.~\eqref{eq: PB}].

The average entropy production is given by
\begin{equation}
\label{eq:appsig}
    \langle \sigma \rangle = \beta \omega \left( \epsilon - p_1 \right),
\end{equation}
implying that work can be extracted as long as the measurement error is small enough, $\epsilon<p_1$. The average coarse-grained entropy reads
\begin{equation}
    \langle \sigma_{\rm cg} \rangle = -H\left(P[0] \right) - \ln{\left(  (1- \epsilon) p_0 + \epsilon p_1 \right)},
\end{equation}
where $H(x) \equiv -x \ln{x} - (1-x) \ln{(1-x)}$. The ensemble average of the mutual information~\cite{Sagawa_2008, Sagawa_2010}, which provides another second law of information thermodynamics [cf. Eq.~\eqref{eq: QC te} for $M = 1$], is given by the expression
\begin{equation}
    \langle I_{\rm mi} \rangle_{\rm c} \equiv S(\op{\rho}_0) - \sum_Y P[Y] S(\op{\rho}^Y) = -H(\epsilon) + H\left(P[0]\right).
\end{equation}
The extracted work, which we define as the average change in the energy resulting from the feedback unitary
\begin{equation}
\label{eq: W calculation}
    \langle W \rangle = \sum_Y P[Y] \left( \text{Tr} \{ \op{H} \op{\rho}^Y  \} - \text{Tr} \{ \op{H} \op{U}_Y \op{\rho}^Y \op{U}_Y^\dagger  \} \right),
\end{equation}
coincides with $-\langle \sigma \rangle/\beta$. This is due to the fact the the measurement Kraus operators commute with the Hamiltonian and do not change the internal energy of the system.

Calculations for the quantum protocol are performed in a similar way, with the same initial state given in Eq.~\eqref{eq:instatesm}. The probabilities of the measurement outcomes are given by $P[+] = P[-] = 1/2$. The post-measured quantum states, conditioned on the outcomes are given by
\begin{equation}
\label{eq: post measurement quantum}
    \begin{split}
        \op{\rho}^+ =&  (1-\epsilon) \proj{+} + \epsilon \proj{-} + \sqrt{\epsilon(1-\epsilon)} (p_0 - p_1) \left( |+\rangle \langle -| + |-\rangle \langle +| \right), \\
        \op{\rho}^- =&  (1-\epsilon) \proj{-} + \epsilon \proj{+} + \sqrt{\epsilon (1-\epsilon)} (p_0 - p_1) \left( |+\rangle \langle -| + |-\rangle \langle +| \right). \\
    \end{split}
\end{equation}
Application of the feedback results in
\begin{equation}
        \op{U}_+ \op{\rho}^+ \op{U}_+^{\dagger} =\op{U}_- \op{\rho}^-  \op{U}_-^{\dagger} =  (1-\epsilon) \proj{0} + \epsilon \proj{1} + \sqrt{\epsilon (1-\epsilon)} (p_0 - p_1) \left( |0\rangle \langle 1| + |1\rangle \langle 0| \right).
\end{equation}
Following the feedback, the system relaxes to $\op{\rho}_\tau = \op{\rho}_0$. 

In the backward experiment, the probability of obtaining the outcome $Y$, given that $\op{U}_Y^{\dagger}$ was applied to the state $\op{\rho}_\tau$ is given by
\begin{equation}
    P_{\rm tr}[-|\op{U}_-^{\dagger}] = P_{\rm tr}[+|\op{U}_+^{\dagger}] = (1- \epsilon) p_0 + \epsilon p_1.
\end{equation}
The coarse-grained entropy can be obtained with Eq.~\eqref{eq: cg calculation}. Its average is given by the expression
\begin{equation}
    \langle \sigma_{\rm cg} \rangle = \ln{(1/2)} - \ln{\left(  (1- \epsilon) p_0 + \epsilon p_1\right)},
\end{equation}
whereas the quantum-classical mutual information~\cite{Sagawa_2008} reads [cf. Eq.~\eqref{eq: QC te} for $M = 1$]
\begin{equation}
    \langle I_{\rm mi} \rangle_{\rm q} \equiv S(\op{\rho}_0) - \sum_Y P[Y] S(\op{\rho}^Y) = H(p_0) - H\left( \frac{1+ \sqrt{1 - 4 \epsilon (1-\epsilon)(1 - (p_0 - p_1)^2)}}{2} \right).
\end{equation}

\begin{table}[h]
    \centering
        \begin{tabular}{|c | c| c | c | c | c | c|| |c | c | c | c | c | c |c|} 
 \hline
 $a$ & $Y$ & $f$ & $q$ & $P[\Gamma]$ & $e^{-\sigma}$ & $P_{\rm tr}[\bar{\Gamma} |\op{U}_Y^\dagger]$ & $a$ & $Y$ & $f$ & q & $ P[\Gamma]$ & $e^{-\sigma}$ & $P_{\rm tr}[\bar{\Gamma} |\op{U}_Y^\dagger]$   \\  
 \hline\hline
 0 & + & 0 & 0 & $p_0 (1-\epsilon) p_0 /2$ & $1$ & $p_0 (1-\epsilon) p_0 /2$ &  0 & + & 0 & $\omega$ & $p_0 \epsilon p_0/2$ & $e^{-\beta \omega}$ & $p_0 \epsilon p_1/2$ \\ 
 \hline
 1 & + & 0 & 0 & $p_1 (1-\epsilon) p_0 /2$  & $\frac{p_0}{p_1}$ & $p_0 (1-\epsilon) p_0 /2$ & 1 & + & 0 & $\omega$ & $p_1 \epsilon p_0 /2$  & $\frac{p_0}{p_1} e^{-\beta \omega}$ & $p_0 \epsilon p_1/2$ \\ 
 \hline
 0 & + & 1 & $-\omega$ & $p_0 (1-\epsilon) p_1/2$ & $\frac{p_1}{p_0} e^{\beta \omega}$ & $p_1 (1-\epsilon) p_0 /2$ & 0 & + & 1 & 0 & $p_0 \epsilon p_1 /2$ & $\frac{p_1}{p_0} $ & $p_1 \epsilon p_1/2$ \\ 
 \hline
 1 & + & 1 & $-\omega$ & $p_1 (1-\epsilon) p_1/2$ & $ e^{\beta \omega} $ & $p_1 (1-\epsilon) p_0 /2$ & 1 & + & 1 & 0 & $p_1 \epsilon  p_1 /2$ & $ 1 $ &$p_1 \epsilon p_1/2$ \\
 \hline
 0 & - & 0 & 0 & $p_0 (1-\epsilon) p_0/2$ & $1$ & $p_0 (1-\epsilon) p_0 /2$ & 0 & - & 0 & $\omega$ & $p_0 \epsilon p_0/2$ & $e^{-\beta \omega}$ & $p_0 \epsilon p_1 /2$ \\ 
 \hline
 1 & - & 0 & 0 & $p_1 (1-\epsilon) p_0/2$  & $\frac{p_0}{p_1} $ & $p_0 (1-\epsilon) p_0 /2$ & 1 & - & 0 & $\omega$ & $p_1 \epsilon p_0/2$  & $\frac{p_0}{p_1} e^{-\beta \omega} $ & $p_0 \epsilon p_1 /2$ \\ 
 \hline
 0 & - & 1 & $-\omega$ & $p_0 (1-\epsilon)  p_1/2$ & $\frac{p_1}{p_0} e^{\beta \omega}$ & $p_1 (1-\epsilon) p_0 /2$ & 0 & - & 1 & 0 & $p_0 \epsilon p_1/2$ & $\frac{p_1}{p_0}$ & $p_1 \epsilon p_1 /2$ \\ 
 \hline
 1 & - & 1 & $-\omega$ & $p_1 (1-\epsilon)  p_1/2$ & $e^{\beta \omega}$ & $p_1 (1-\epsilon) p_0 /2$ & 1 & - & 1 & 0 & $p_1 \epsilon p_1/2$ & 1 & $p_1 \epsilon p_1/2$ \\
 \hline
\end{tabular}
    \caption{The qubit model with a single quantum measurement. Table shows all the combinations of $\Gamma = \{a, Y, f, q\}$, probability $P[\Gamma]$, corresponding stochastic entropy production $\sigma$, and probability $P_{\rm tr}[\bar{\Gamma} |\op{U}_Y^\dagger]$. Here $q$ in $\Gamma$ corresponds to $-q$ in $\bar{\Gamma}$.}
    \label{tab: quant}
\end{table}

Similarly to the classical case, we group together all quantum jump trajectories with the same net stochastic heat $q = \omega, 0, -\omega$. After this grouping, a trajectory is then defined by $\Gamma = \{a, Y, f, q\}$. Note that in contrast to the classical case, $q$ is not uniquely identified by $a$, $Y$, and $f$. In Tab.~\ref{tab: quant},  we list all different trajectories providing the corresponding probabilities and entropy productions. The probability for a trajectory is given by
\begin{equation}
    P[\Gamma] = p_ap_f |\langle b|\hat{U}_Y\hat{M}_Y|a\rangle|^2= \frac{p_ap_f}{2}\left((\delta_{f,0}\delta_{q,0}+\delta_{f,1}\delta_{q,-\omega})(1-\epsilon)+(\delta_{f,1}\delta_{q,0}+\delta_{f,0}\delta_{q,\omega})\epsilon\right),%p_{a =1} p_{f=0} \text{Tr} \left\{ \op{M} \proj{a}\right\}
\end{equation}
where $b=q/\omega+f$.
From Tab.~\ref{tab: quant}, one may verify $\langle e^{-(\sigma - \sigma_{\rm cg})}\rangle = 1$ For the ensemble average of the entropy production we find the same value as for the classical model, given in Eq.~\eqref{eq:appsig}.
Unlike in the classical case, the extracted work [see Eq.~\eqref{eq: W calculation}] is not equal to $-\langle \sigma \rangle/\beta$. Instead, we obtain
\begin{equation}
    \langle W \rangle = \left( \frac{1}{2} - \epsilon \right) \omega.
\end{equation}
The discrepancy between $-\langle \sigma \rangle$ and $\beta \langle W \rangle$ arises due to noncommutativity of the measurement Kraus operators and the Hamiltonian. The measurement changes the energy of the quantum state, an effect that has been referred to as quantum heat~\cite{Elouard_2017_qh, Elouard_2017_engine}. Finally, the backward probability can be found from $P_{\rm B}[\bar{\Gamma}] = P_{\rm tr}[\bar{\Gamma} | \op{U}_Y^\dagger ] P[Y]/P_{\rm tr}[Y | \op{U}_Y^\dagger ]$ [cf. Eq.~\eqref{eq: PB}], where
\begin{equation}
    P_{\rm tr}[\bar{\Gamma} | \op{U}_Y^\dagger ] = p_f \text{Tr}\langle b|\op{\rho}_\tau|b\rangle |\langle b|\hat{U}_Y\hat{M}_Y|a\rangle|^2 = P[\Gamma]\frac{\langle b|\op{\rho}_\tau|b\rangle}{p_a}, 
\end{equation}
and $b = f + q/\omega$.

\subsection{Two measurements}

We now turn to the scenario, where measurement and feedback are performed twice. The protocol is similar to the single measurement case. First, a measurement with Kraus operator $\hat{M}_{y_1}$ is performed. Then, a unitary depending on the measurement outcome, $\hat{U}_{y_1}$. is applied. After this first feedback round, the system is interacting with a thermal bath for a time $\Delta t$. This is modeled by the Lindblad master equation described in the main text with Liouvillian $\mathcal{L}$. After this partial thermalization, another round of measurement ($\hat{M}_{y_2}$) and feedback ($\hat{U}_{y_2}$) is applied. Finally, the system thermalizes, returning to the thermal state.

In this case, we describe trajectories by the variables $\Gamma=\{a,y_1,y_2,f,q_1,q_2\}$, where $q_j$ denotes the heat exchanged with the environment after measurement $j$. The probabilities can be calculated similarly to the single measurement case
\begin{equation}
\begin{aligned}
    &P[\Gamma|_{q_1=\omega}] 
    = p_ap_f|\langle b|\hat{U}_{y_2}\hat{M}_{y_2}|0\rangle|^2|\langle 1|\hat{U}_{y_1}\hat{M}_{y_1}|a\rangle|^2\langle 0|e^{\mathcal{L}\Delta t}\left\{\proj{1}\right\}|0\rangle,\\
    &P[\Gamma|_{q_1=-\omega}] 
    = p_ap_f|\langle b|\hat{U}_{y_2}\hat{M}_{y_2}|1\rangle|^2|\langle 0|\hat{U}_{y_1}\hat{M}_{y_1}|a\rangle|^2\langle 1|e^{\mathcal{L}\Delta t}\left\{\proj{0}\right\}|1\rangle,\\
    &P[\Gamma|_{q_1=0}] 
    = p_ap_f\sum_{c,d=0,1}\langle b|\hat{U}_{y_2}\hat{M}_{y_2}|c\rangle\langle c|\hat{U}_{y_1}\hat{M}_{y_1}|a\rangle\langle a|\hat{M}^\dagger_{y_1}\hat{U}^\dagger_{y_1}|d\rangle\langle c|e^{\mathcal{L}\Delta t}\left\{|c\rangle\langle d|\right\}|d\rangle\langle d|\hat{M}^\dagger_{y_2}\hat{U}^\dagger_{y_2}|b\rangle,
    \end{aligned}
\end{equation}
where again $b=q_2/\omega +f$. In total, this results in 96 possible trajectories. We refrain from providing a table with all probabilities. They can be evaluated in a straightforward manner using the identities:
\begin{equation}
    \begin{split}
        & \bra{\ell'} \op{U}_y \op{M}_y \ket{\ell} = \left( \delta_{y, \ell} \sqrt{1-\epsilon} + (1-\delta_{y, \ell}) \sqrt{\epsilon} \right) \left( \delta_{y, 0} \delta_{\ell', \ell} + \delta_{y, 1} (1-\delta_{\ell', \ell} ) \right) \text{     (classical case)}\\
        & \bra{\ell'} \op{U}_y \op{M}_y \ket{\ell} = \left( \delta_{y, +} + \delta_{y, -} (-1)^\ell  \right) \left( \delta_{\ell', 0} \sqrt{(1-\epsilon)/2} +(-1)^\ell \delta_{\ell', 1} \sqrt{\epsilon/2} \right) \text{     (quantum case)}\\
        & \bra{0} e^{\mc{L} \Delta t} \left\{ \proj{0} \right\} \ket{0}= \frac{   e^{-\beta \omega} e^{-\kappa(1+ e^{-\beta \omega}) t} + 1 }{
        1+ e^{-\beta \omega}} \\
        & \bra{1} e^{\mc{L} \Delta t} \left\{ \proj{0} \right\} \ket{1}=  \frac{   e^{-\beta \omega} \left( 1- e^{-\kappa(1+ e^{-\beta \omega}) t} \right) }{
        1+ e^{-\beta \omega}} \\
        & \bra{0} e^{\mc{L} \Delta t} \left\{ \proj{1} \right\} \ket{0}=  \frac{   \left( 1- e^{-\kappa(1+ e^{-\beta \omega}) t} \right) }{
        1+ e^{-\beta \omega}} \\
        & \bra{1} e^{\mc{L} \Delta t} \left\{ \proj{1} \right\} \ket{1}= \frac{   e^{-\kappa(1+ e^{-\beta \omega}) t} + e^{-\beta \omega} }{
        1+ e^{-\beta \omega}}\\
        & \bra{0} e^{\mc{L} \Delta t} \left\{ \ket{0}\bra{1} \right\} \ket{1} = e^{i \omega t} e^{- \kappa (1+ e^{-\beta \omega})/2}  \\
        & \bra{1} e^{\mc{L} \Delta t} \left\{ \ket{1}\bra{0} \right\} \ket{0} = e^{i \omega t} e^{- \kappa (1+ e^{-\beta \omega})/2},\\
    \end{split}
\end{equation}
where $\ell, \ell' = 0, 1$. The probability for obtaining the measurement outcomes $Y = \{y_2, y_1\}$ can be found from the expression
\begin{equation}
    P[Y] = \text{Tr} \left\{ \op{M}_{y_2} e^{\mc{L} \Delta t} \left\{ \op{U}_{y_1} \op{M}_{y_1} \op{\rho}_0 \op{M}_{y_1}^\dagger \op{U}_{y_1}^\dagger  \right\} \op{M}_{y_2}^\dagger \right\}.
\end{equation}
Similarly, the probability of obtaining $\bar{Y} = \{y_1, y_2\}$ when we reverse the feedback unitary operations is given by
\begin{equation}
    P_{\rm tr}[\bar{Y}| U_Y^\dagger] = \text{Tr} \left\{ \op{M}_{y_2}^\dagger e^{\mc{L} \Delta t} \left\{ \op{U}_{y_2}^\dagger \op{M}_{y_2}^\dagger \op{\rho}_\tau \op{M}_{y_2} \op{U}_{y_2}  \right\} \op{M}_{y_1} \right\}.
\end{equation}
These quantities allow to compute the coarse-grained entropy $\sigma_{\rm cg} [Y]$. Our calculations are consistent with our integral FT $\langle e^{-(\sigma - \sigma_[\rm cg])} \rangle =1$. Full backward probability $P_{\rm B}[\bar{\Gamma}]$ [cf. Eq.~\eqref{eq: PB}] can be obtained using
\begin{equation}
\begin{aligned}
    &P[\bar{\Gamma}|_{q_1=\omega}| U_Y^\dagger] 
    = p_f \bra{b} \op{\rho}_\tau \ket{b} |\langle b|\hat{U}_{y_2}\hat{M}_{y_2}|0\rangle|^2 
        \langle 1|e^{\mathcal{L}\Delta t}\left\{\proj{0}\right\}|1\rangle
        |\langle 1|\hat{U}_{y_1}\hat{M}_{y_1}|a\rangle|^2,\\
    &P[\bar{\Gamma}|_{q_1=-\omega}| U_Y^\dagger] 
    = p_f \bra{b} \op{\rho}_\tau \ket{b} |\langle b|\hat{U}_{y_2}\hat{M}_{y_2}|1\rangle|^2 
        \langle 0|e^{\mathcal{L}\Delta t}\left\{\proj{1}\right\}|0\rangle
        |\langle 0|\hat{U}_{y_1}\hat{M}_{y_1}|a\rangle|^2,\\
    &P[\bar{\Gamma}|_{q_1=0}| U_Y^\dagger] 
    =  p_f \bra{b} \op{\rho}_\tau \ket{b} \sum_{c, d = 0, 1} \bra{a} \hat{M}_{y_1}^\dagger \hat{U}_{y_1}^\dagger \ket{c} \bra{c} \hat{M}_{y_2}^\dagger \hat{U}_{y_2}^\dagger \proj{b} \hat{U}_{y_2} \hat{M}_{y_2} \ket{d} \bra{d} \hat{U}_{y_2} \hat{M}_{y_2} \ket{a}
        \langle c| e^{\mathcal{L}\Delta t}\left\{\ket{c} \bra{d} \right\}|d\rangle,
    \end{aligned}
\end{equation}
where $b = f+ q_2/\omega$. The average extracted work $\langle W \rangle$, which is given by the generalization of Eq.~\eqref{eq: W calculation} to the two measurements case, is equal to $-\langle \sigma \rangle/\beta  $ in the classical case. In the quantum case, these two quantities cannot be equated, which was already discussed for the single measurement scenario.

The (quantum-classical) transfer entropy $\langle I_{\rm te} \rangle$, which givens another bound on the entropy production, can be obtained using Eq.~\eqref{eq: QC te} with $M = 2$. In this expression, the density matrix $\op{\rho}_{t_1}^{Y_0}$ corresponds to $\op{\rho}_0$. Similarly, $\op{\rho}_{t_1}^{Y_1}$ corresponds to $\op{\rho}^{y_1}$ [see Eqs.~\eqref{eq: post measurement classical} and~\eqref{eq: post measurement quantum}], $\op{\rho}_{t_2}^{Y_1}$ corresponds to $e^{\mc{L} t}\left\{ \op{U}_{y_1} \op{\rho}^{y_1} \op{U}_{y_1}^\dagger \right\}/ \text{Tr} \left\{ e^{\mc{L} t}\left\{ \op{U}_{y_1} \op{\rho}^{y_1} \op{U}_{y_1}^\dagger \right\} \right\}$, and 
$\op{\rho}_{t_2}^{Y_2}$ corresponds to $\op{M}_{y_2} \op{\rho}_{t_2}^{Y_1} \op{M}_{y_2}^\dagger / \text{Tr} \left\{ \op{M}_{y_2} \op{\rho}_{t_2}^{Y_1} \op{M}_{y_2}^\dagger \right\}$.

\section{V. Continuous measurement}

Continuous measurements of two types may be considered: First, we may consider the limit where measurements are performed repeatedly, with a fixed time $\delta t$ in between measurements. Then, the limit $\delta t\rightarrow 0$ is taken. To avoid the Zeno effect, the strength of the measurement is usually taken to go to zero at the same time, such that the rate of acquiring information remains finite. The POVM elements are usually taken to be Gaussian, which can be motivated from the central limit theorem. See Ref.~\cite{jacobs_2006} for more information for this type of continuous measurement. Our FT holds for this type of continuous measurements as they are a special case of the multiple measurement scenario discussed in Secs.~I and II above.

Alternatively, we may continuously monitor the jumps in a Lindblad master equation \cite{Yada_2022,landi_2023}. In this section, we focus on this type of measurement. The time-evolution of the system may then be described by a Lindblad master equation, where feedback is applied depending on the observed jumps. This scenario may be described by an unravelled master equation that describes the evolution of the conditional quantum state 
\begin{equation}
\label{eq: Stoch ME cont}
\begin{split}
    d\op{\rho}_{\rm c} =& -i \left( \op{H}_{\rm eff}(\lambda_{t}^{Y_t}) \op{\rho}_{\rm c} - \op{\rho}_{\rm c} \op{H}_{\rm eff}^\dagger(\lambda_{t}^{Y_t})  \right) d t  + \sum_j \text{Tr} \{ \op{L}_j^{\dagger}(\lambda_t^{Y_t}) \op{L}_j(\lambda_t^{Y_t}) \op{\rho}_{\rm c} \}  \op{\rho}_{\rm c} dt  + \sum_y \text{Tr}\{ \op{M}_y^\dagger\op{M}_{y} \op{\rho}_{\rm c}  \}\op{\rho}_{\rm c} dt  \\
    +&  \sum_j \left( \frac{ \op{L}_j(\lambda_t^{Y_t}) \op{\rho}_{\rm c} \op{L}_j^{\dagger}(\lambda_t^{Y_t}) }{\text{Tr} \{ \op{L}_j^{\dagger}(\lambda_t^{Y_t}) \op{L}_j(\lambda_t^{Y_t}) \op{\rho}_{\rm c} \} }- \op{\rho}_{\rm c} \right) dN_j  + \sum_y  \left( \frac{\op{V}_y({Y_{t}})\op{M}_y \op{\rho}_{\rm c} \op{M}_y^{\dagger}\op{V}_y^\dagger(Y_{t})}{\text{Tr} \{ \op{M}^{\dagger}_y \op{M}_y \op{\rho}_{\rm c} \}}- \op{\rho}_{\rm c} \right) dM_y,
\end{split}
\end{equation}
where we introduced two types of jumps: those which are observed, described by the operators $\hat{M}_y$, and those that are not observed, described by the operators $\hat{L}_j$.
The random variable $d N_j$ ($dM_y$) takes the value $1$ if the jump $j$ (measurement jump with outcome $y$) occurs and $0$ otherwise. These random variables obey $d M_y d M_{y'} = d M_y \delta_{y, y'}$,  $d N_j d N_{j'} = d N_j \delta_{j, j'}$, and $dM_y dN_j = 0$. Here, the effective non-Hermitian Hamiltonian is given by
\begin{equation}
\label{eq: NH cont}
        \op{H}_{\rm eff}(\lambda_t^{Y_t}) = \op{H}(\lambda_t^{Y_t}) - \frac{i}{2} \sum_j \op{L}_j^{\dagger}(\lambda_t^{Y_t}) \op{L}_j(\lambda_t^{Y_t}) - \frac{i}{2} \sum_y \op{M}_y^{\dagger} \op{M}_y.
\end{equation}
Quantum jump detections now happen at random times $t_n$, resulting in outcomes $y_n$. The measurement outcomes up to time $t$ can then be described by all jumps that happen before this time, i.e., $Y_t=\{(t_1,y_1),(t_2,y_2),\cdots (t_n,y_n)|t_n<t\leq t_{n+1}\}$. As above, feedback is mediated by the protocol $\lambda_t(Y_{t})$, which may alter both the Hamiltonian and the unobserved jump operators. In addition, a unitary $\hat{V}_{y_n}(Y_{t_n})$ may be applied directly after observing the jump at time $t_n$ corresponding to outcome $y_n$. In addition to the very last outcome, this unitary may also depend on the previous measurement outcomes through $Y_{t_n}$.

%The driving parameter $\lambda_t^{Y_n}$ depends on the $n$ jump detections that occurred prior to the time $t$. Upon observation of $y_{n+1}$ we update $\lambda_t^{Y_n}$ to $\lambda_t^{Y_{n+1}}$. In addition, we may apply a feedback pulse $\op{V}^{Y_{n+1}}$. This type of measurement was used by the authors of Ref.~\cite{Yada_2022} to derive the FT and the second law with the quantum-classical transfer entropy $I_{\rm te}$. 

In this framework, the trajectory in the forward experiment may be defined as $\Gamma = \{ Y, \gamma\}$, where $\gamma$ is defined analogously to the quantum jump unravelling of the Lindblad master equation, and $Y \equiv Y_\tau$ where (as above), $\tau$ denotes the total time of the trajectory. The probability density of the trajectory can then be expressed as
\begin{equation}
\label{eq: P Gamma cont}
    P[\Gamma] = p_a \text{Tr} \left\{ \proj{f_Y}  \mc{T} \left\{  \mc{U}_{\rm eff}(\tau,0) \prod_{n = 1}^M \mc{M}_n(y_n,t_n) \prod_{k = 1}^K \mc{J}_k(j_k, s_k) \right\}   \proj{a}  \right\}  ,
\end{equation}
where 
\begin{equation}
    \label{eq:jumpsuper}
    \mc{M}_n(y_n,t_n)\hat{\rho} = \op{V}_{y_n}({Y_{t_n}})\hat{M}_{y_n}\hat{\rho}\hat{M}_{y_n}^\dagger\op{V}^\dagger_{y_n}({Y_{t_n}}),
\end{equation}
$\mc{J}_k(j_k, s_k)$ is given by Eq.~\eqref{eq: Jump}, $\mc{U}_{\rm eff}(\tau,0)$ by Eqs.~\eqref{eq: U super eff} and \eqref{eq: U eff} with the non-Hermitian Hamiltonian in Eq.~\eqref{eq: NH cont}.
%we have used the notation of Eq.~\eqref{eq: P Gamma Lindblad}. Eq.~\eqref{eq: U eff} now incorporates non-Hermitian Hamiltonian in Eq.~\eqref{eq: NH cont} instead of one in Eq.~\eqref{eq: NH}, and we substitute $\op{M}_n(y_n) \to \op{V}^{Y_n} \op{M}_n(y_n)$ in Eq~\eqref{eq: Measurment} to account for the feedback pulse. 
We note that Eqs.~\eqref{eq: P Gamma Lindblad} and~\eqref{eq: P Gamma cont} have the same form. In the backward experiment, the probability density of the inverse trajectory is given by
\begin{equation}
\label{eq: P Gamma cont tr}
    P_{\rm tr}[\bar{\Gamma}|\{\lambda_{\tau -t}^Y \}] = p_{f_Y} \text{Tr} \left\{ \op{\Theta} \proj{a} \op{\Theta}^{-1} \mc{T} \left\{ \bar{\mc{U}}_{\rm eff}(\tau, 0) \prod_{n = 1}^M \bar{\mathcal{M}}_{n}(y_n, \tau - t_n)  \prod_{k = 1}^K \bar{\mc{J}}_k(\bar{s}_k, \bar{j}_k) \right\}   \op{\Theta} \proj{f_Y} \op{\Theta}^{-1}  \right\},
\end{equation}
where
\begin{equation}
    \bar{\mc{M}}_n(y_n,\tau-t_n)\hat{\rho} = \op{\Theta} \op{M}^\dagger_{y_n} \op{V}_{y_n}^\dagger({Y_{t_n}}) \op{\Theta}^{-1} \hat{\rho} \op{\Theta} \op{V}_{y_n}({Y_{ t_n}})  \op{M}_{y_n}  \op{\Theta}^{-1} 
\end{equation}
is applied at the time $\tau - t_n$, $\bar{\mc{J}}_k(\bar{s}_k, \bar{j}_k)$ is given by Eq.~\eqref{eq: Jump B}, and $\bar{\mc{U}}_{\rm eff}$ describes time-evolution with the non-Hermitian Hamiltonian is given by $\op{\Theta} \op{H}_{\rm eff}^\dagger(\lambda^Y_{\tau-t}) \op{\Theta}^{-1}$ [cf. Eq.~\eqref{eq: NH cont}], which as above ensures a microreversibility condition. Eq.~\eqref{eq: P Gamma cont tr} has again the same form as Eq.~\eqref{eq: P tr Lindblad}. These imply that the detailed FT [see Eq.~\eqref{eq: FT sigma}] holds, where the stochastic entropy production $\sigma[\Gamma]$ is given by Eq.~\eqref{eq: sigma Lindblad} as in the case of the Lindblad master equation. Likewise, for a given $Y$, the derivation of Eq.~\eqref{eq: sigma cg m} for the coarse-grained entropy follows the same lines as in the case of a Lindblad master equation with discrete measurements. The integral FT follows from a calculation analogous to Eq.~\eqref{eq: Int FT derivation Lind}, where the summation over $Y$ now involves summation over all $M$ from $0$ to $\infty$ as well as integration over all possible times of the measurement.
%\begin{equation}
%\begin{split}
%     \langle e^{-(\sigma - \sigma_{\rm cg})} \rangle =& \sum_Y \sum_\gamma P[\Gamma] e^{-(\sigma[\Gamma] - \sigma_{\rm cg}[Y])}  =  \sum_{Y} \sum_{\bar{\gamma}} P_{\rm B}[\bar{\Gamma}|\{\lambda_{\tau - t}^Y \}] \\
%     = & \sum_{Y} \frac{P[Y]}{P_{\rm tr}[\bar{Y}|\{\lambda_{\tau - t}^Y \}]} \int \mathfrak{D}[\bar{\gamma}] P_{\rm tr}[\bar{\Gamma}|\{\lambda_{\tau - t}^Y \}] =   \sum_{Y} \frac{P[Y]}{P_{\rm tr}[\bar{Y}|\{\lambda_{\tau - t}^Y \}]}  P_{\rm tr}[\bar{Y}|\{\lambda_{\tau - t}^Y \}] = 1,
%\end{split}
%\end{equation}

Our results for the time evolution described in Eq.~\eqref{eq: Stoch ME cont} could be alternatively derived by dividing the duration of the experiment $\tau$ into intervals $\delta t$. At each time point $t_n = n \delta t$, either a jump $j$ happens with probability $\text{Tr} \{\op{L}_j^\dagger \op{L}_j \op{\rho}_{\rm c} \} \delta t$, or measurement jump $y$ happens with the probability $\text{Tr} \{\op{M}_j^\dagger \op{M}_j \op{\rho}_{\rm c} \} \delta t$, or neither of them occur. The latter corresponds to application of the Kraus operator $\op{I} - i \delta t \op{H}_{\rm eff}$ [cf. Eq.~\eqref{eq: NH cont}]. We note that in order to obtain the quantum-classical transfer entropy $\langle I_{\rm te} \rangle$ [cf. Eq.~\eqref{eq: QC te}], it is necessary to use an unraveling where jumps $j$ and jumps $y$ occur sequentially in each time step, thus allowing for both types of jumps in a single interval $\delta t$ (see Ref.~\cite{Yada_2022} for details). This approach is equivalent to the standard unravelling up to the order of $\delta t$, and could be also applied to derive our results. For the details about computing $\langle I_{\rm te} \rangle$ we refer the reader to Ref.~\cite{Yada_2022}.

\subsection{Qubit Model}

The Hamiltonian of the system is given by $\op{H}_t = \frac{\omega}{2} \op{\sigma}_z +  \chi \cos{(\Omega t)}$. The jump operators $\op{L}_- = \kappa \op{\sigma}_-$ and $\op{L}_+ = e^{-\beta \omega}\kappa \op{\sigma}_+$ are ladder operators of $ \frac{\omega}{2} \op{\sigma}_z$. The continuous measurement is performed by monitoring a single jump operator $\op{M}  = \sqrt{\kappa_{\rm m} } (\sqrt{1-\epsilon} \proj{1} + \sqrt{\epsilon} \proj{0})$ corresponding to detection of the excited state with error probability $\epsilon$. %and a Kraus operator $\op{I} - \frac{ \kappa_{\rm m} \delta t }{2} \left( (1- \epsilon) \proj{1} + \epsilon \proj{0} \right) $ corresponding to no detection. 
When a jump is observed, a unitary rotation $\op{\sigma}_x$ is applied to bring the system to the ground state. The numerical results are performed using a quantum jump simulation, where the duration of the protocol is divided into $\tau/\delta{t} = 1000$ steps. The number of randomly sampled trajectories for each data point in Fig.~\ref{fig:ContinuousToyModel} is $10^5$.

%In Ref.~\cite{Yada_2022} the second law of information thermodynamics $\langle \sigma \rangle \geq - \langle I_{\rm te} \rangle $ was introduced for continuous measurement and feedback of open quantum systems governed by Eq.~\eqref{eq: Stoch ME cont}. In this framework, the quantum-classical transfer entropy, which bounds the entropy production, is given by
%\begin{equation}
%   \langle I_{\rm te} \rangle = \sum_{n = 0}^{M-1} \sum_{Y_n} P[Y_n] \mathcal{I}\left( \op{\rho}_{t_n}^{Y_n} , \mathcal{M}_{n+1} (y_{n+1}) \right),
%\end{equation}
%where
%\begin{equation}
%    \mathcal{I}\left( \op{\rho}_{t_n}^{Y_n} , \mathcal{M}_{n+1} (y_{n+1}) \right) = S_{\rm vn}(\op{\rho}_{t_n}^{Y_n}) - \sum_{y_{n+1}} P[y_{n+1}|Y_n ] S_{\rm vn}\left( \frac{ \mathcal{M}_{n+1} (y_{n+1}) \op{\rho}_{t_n}^{Y_n}}{P[y_{n+1}|Y_n ]} \right) 
%\end{equation}\
%and $P[y_{n+1}|Y_n ] = \text{Tr} \{ \mathcal{M}_{n+1} (y_{n+1}) \op{\rho}_{t_n}^{Y_n}\}$.

\section{VI. Time-reversed measurement}

The POVM measurement in the forward protocol is described with the set of the Kraus operators $\{ \mathcal{M}(y) \}$. The simplest kind of the measurement is a projective measurement, when the Kraus operators fulfil $\mathcal{M}(y) = \mathcal{M}^\dagger(y)$ and $\mathcal{M}(y) \mathcal{M}(y') = \delta_{y, y'} \mathcal{M}(y)$. In this case, in the backward measurement we simply perform the same measurement as in the forward protocol, with the set of the Kraus operators $\{ \op{\Theta} \mathcal{M}(y) \op{\Theta}^{-1}  \}$. If the measurement is not projective, but it is unital, i.e., it obeys $\sum_y \mathcal{M}(y) \mathcal{M}^{\dagger}(y) = \op{I}$ (not to be confused with $\sum_y \mathcal{M}^{\dagger}(y) \mathcal{M}(y) = \op{I}$ which is obeyed for all measurements), then in the backward experiment we perform a measurement with the set of the Kraus operators $\{ \op{\Theta} \mathcal{M}^{\dagger}(y) \op{\Theta}^{-1} \}$.

In the general case, when the measurement is neither projective, nor unital, we follow~\cite{Gong_2016}. The action of each Kraus operator can be modeled as a projective measurement on an ancilla A that is correlated with the measured system S. Suppose that before the measurement, the ancilla is in the state $\proj{\alpha}$ and the system in $\op{\rho}$. The measurement with the outcome $y$ consists of correlating the ancilla with the system using the joint unitary $U$ and collapsing the state of the ancilla onto $\proj{\alpha_y}$. The corresponding Kraus operator is given by $ \mathcal{M}(y) = \bra{\alpha_y}  U  \ket{\alpha}$. In the backward measurement, the Kraus operator $\op{\Theta} \mathcal{M}^\dagger(y) \op{\Theta}^{-1}$ results from preparing the ancilla in the state $\op{\Theta} \proj{\alpha_y} \op{\Theta}^{-1}$, acting with $\op{\Theta} U^\dagger \op{\Theta}^{-1} $ onto the joint state of the ancilla and the state, and making a projective measurement onto the ancilla, such that its state collapses to $\op{\Theta} \proj{\alpha} \op{\Theta}^{-1}$. Thus, in the backward experiment $y$ is no longer a measurement outcome. Instead, for each $y$ a different measurement is performed and only one of the outcomes (resulting in the ancilla being in the state $\alpha$) is relevant. For any other outcome, the result needs to be discarded. In our FT, this does not result in an additional overhead since in any case, measurement outcomes that do not agree with what is observed in the corresponding forward trajectory need to be discarded.

%Using the Schmidt orthogonalisation, we can always choose an $M$ dimensional basis of the ancilla $\{ \tilde{\alpha}_n \}$ fulfilling $\sum_{n=0}^{M-1} \proj{\tilde{\alpha}_n} = \op{I}_{\rm A}$ and $\langle \tilde{\alpha}_n | \tilde{\alpha}_{n'} \rangle = \delta_{n, n'}$, such that $\proj{\tilde{\alpha}_0} = \proj{\alpha}$. Therefore, preparing the state of the ancilla $\proj{\alpha_y}$, applying $U^\dagger$, and making a projective measurement of the ancilla with $\{ \tilde{\alpha}_n \}$, gives rise to the POVM measurement, where one of its Kraus operators is equal to $\mathcal{M}^\dagger(y)$.  

\end{document}